\documentclass[aps,prl,twocolumn,showpacs,preprintnumbers]{revtex4-1}

\usepackage{psfrag,graphicx}
\usepackage{dcolumn}
\usepackage{amsmath,amssymb}
\usepackage{bm}
\usepackage{amsfonts,amssymb,amsmath}        

\newcommand{\be}{\begin{equation}}
\newcommand{\ee}{\end{equation}}
\newcommand{\bq}{\begin{eqnarray}}
\newcommand{\eq}{\end{eqnarray}}

\newcommand{\ket}[1]{\left | \, #1 \right\rangle}

\begin{document}

\title{Quantum memories and error correction}
\author{James R. Wootton}
\affiliation{Department of Physics, University of Basel, Klingelbergstrasse 82, CH-4056 Basel, Switzerland}

\begin{abstract}

Quantum states are inherently fragile, making their storage a major concern for many practical applications and experimental tests of quantum mechanics. The field of quantum memories is concerned with how this storage may be achieved, covering everything from the physical systems best suited to the task to the abstract methods that may be used to increase performance. This review concerns itself with the latter, giving an overview of error correction and self-correction, and how they may be used to achieve fault-tolerant quantum computation. The planar code is presented as a concrete example, both as a quantum memory and as a framework for quantum computation.

\end{abstract}


\maketitle

\section{Introduction}

What is a quantum memory? The answer to this question is straightforward: it is something that can be used to store a quantum state. But a quantum state of what? Stored for what purpose? These questions lead to a multiplicity of answers. And this multiplicity leads to many diverse fields of research throughout quantum optics, quantum information and condensed matter \cite{qkd1,qkd2,qkd3,qrep1,qrep2,linopt1,linopt2,linopt3,network1,network2,toric,dennis}.

Research into quantum memories can be split into two main fields: the \textit{physical} and the \textit{logical}. In reality the split is not so neat with some approaches belonging to both, but it is a useful fiction for the purpose of this review. The field of \textit{physical} quantum memories concerns the actual physical systems that may be used to store a quantum state, systems such as quantum dots, rare-earth ions in solids, NV centres in diamond, Raman memories, cold trapped atomic ensembles and many others. In-depth reviews of such systems can be found in \cite{review1,review2,review3,review4,review5}. 

Each of these approaches has its own strengths and weaknesses. They allow different operations to be performed on the stored information, and different ways to interface the memory with other systems. The physical system used for a quantum memory will therefore depend on the application that the memory is to be used for.

Some applications, such as certain proposals for quantum computation, require quantum memories to store states with very high fidelity for long times. These high requirements can be beyond what we can hope any physical memory to achieve on its own. It will need some help to keep errors in check, to keep the fidelity high and elongate the decoherence time. It's research into how this help may be achieved that is the focus of research into \textit{logical} quantum memories. The typical approach is that of multipartite encoding. Instead of storing a single qubit in a single physical memory, a large number of physical memories are used. The qubit, now called a \textit{logical qubit}, is then stored non-locally across the entire multi-memory system. Any error on one memory can be seen by comparing its state to that of its neighbours. This is done either through measurement, allowing the error to be removed by quantum error correction \cite{coderev}, or by implementing a Hamiltonian to energetically suppress the errors. Many proposals have been put forward for such logical encodings, capable of removing multiple forms of errors including spurious operations on systems as well as loss of some physical memories. A (by no means complete) list of some examples of error correction in simple systems can be found here \cite{example1,example2,example3,example4,example5}.

The restrictions and requirements of the physical memories will affect what sort of encoding is possible for the logical qubits, as well as what sort of encoding is required to best protect against the typical errors of the system. As such, it is important for there to be a deep understanding between those who work in these fields. It is a study of these the logical quantum memories, and their place in quantum computation, that concerns the bulk of this review.  {In particular we consider the capabilities of Kiteav's surface codes, which include the famous toric code, to act as both a quantum memory, and as a fault-tolerant platform for quantum computation} \cite{toric,dennis}.

\section{Why should we store a quantum state?}

First let us consider a few of the applications of a quantum memory, with particular emphasis on those applications that require the use of a logical encoding on top of the physical one.

\subsection{Quantum memories and flying qubits}

Many applications of quantum information theory require the use of so-called flying qubits. These typically take the form of photons, with the qubit state encoded within the photon polarization. Flying qubits allow quantum information to be transferred over potentially long distances, providing the basic resource for many important applications such as quantum communication protocols, experimental probes of quantum foundations and the transfer of quantum information between different parts of a quantum computer.

Even so, it isn't enough just to have photons whizzing around. The qubits need to be processed in order to extract useful information, and undo the effects of errors that occur during their journeys. For this, they need to be caught, stored, manipulated, measured and released in a deterministic manner. These are the tasks for which quantum memories are required. The exact relationship that quantum memories need to have with flying qubits depends on the application. We will now consider a few examples.

\subsubsection{Quantum repeaters}

Many proposed applications of quantum technology rely on the possibility of long-range quantum state transfer, including quantum key distribution \cite{qkd1,qkd2,qkd3} and quantum networks \cite{network1,network2}. Just as long-range transmission of classical information is crucial in present day communications, the equivalent quantum process is likely to become the workhorse of the future. However, transmission of quantum information over long distances will always incur heavy losses. And, unlike in the classical case, copying of the information to amplify the signal is forbidden by the no-cloning theorem \cite{noclone1,noclone2}. As such, a quantum solution must be provided for this quantum problem.

A solution, proposed by Hans Briegel et al. in 1998, is known as the quantum repeater approach \cite{qrep1}. For this, a number of quantum repeater stations are set up between the sender and recipient. The long distance over which the information is to be sent is therefore split up into many smaller links. Each station will then create entangled pairs, keeping one particle of each and sending the other to one of its neighbours. By doing so, neighbouring stations will share entanglement. Entanglement swapping \cite{eswap1,eswap2} and purification \cite{pure} may then be used to extend the range of this entanglement, ultimately providing a number of entangled pairs shared by the sender and recipient. Of course, the noise incurred when particles are sent between neighbouring stations will mean that this entanglement is imperfect. Nevertheless, the use of purification can distil many noisy copies into one of high fidelity. With this, a quantum state may be teleported directly from sender to recipient without error \cite{tele}, even though no particle has had to travel further than the distance between two stations.

Quantum memories are crucial to the implementation of a quantum repeater. Without storage of the entangled states created and received by each station, the entanglement between them cannot be maintained. It also cannot be manipulated to create the purified entanglement that stretches the entire distance of the network, required for the ultimate transmission by teleportation. For these things to be achieved, it must be possible to store the quantum states of the photons sent between neighbouring stations. The quantum memory used for this must also have a long coherence time, and allow the operations required for entanglement swapping and state purification. To achieve this, quantum repeaters may employ quantum error correction \cite{jiang,fowrep,munro}.

\subsubsection{Quantum foundations}

The weight of experimental evidence supports the fact that Bell's inequalities are violated in nature \cite{bell1,bell2}. Nevertheless, all current experiments have suffered from loopholes that still allow interpretations consistent with local realism to be made. It is therefore of fundamental interest to close the gaps as much as possible, to further demonstrate the non-local nature of entanglement \cite{bell3}. Quantum memories help toward this end in two ways. Firstly they aid the preparation of high fidelity entangled pairs separated by the long distances required to ensure locality. This can be done, for example, using the quantum repeater technique described above. Secondly they provide more reliable interfacing with the quantum state, allowing measurements to be made with the high efficiencies required to close the fair sampling loophole.

Another cornerstone of quantum mechanics is the uncertainty principle. However, an entropic reformulation of this has led to a more detailed understanding of exactly what can and cannot be concurrently known. Specifically, it is found that entangling a system to a quantum memory allows measurements to be made that are apparently more accurate than Heisenberg allows \cite{uncer}. Of course this is not the case, and the results will always be in perfect agreement with the generalized uncertainty relationship that takes the entanglement into account, but it demonstrates the power of quantum memories in probing the dark corners of quantum mechanics.

\subsection{Quantum computation}

Quantum computation aims to quickly perform computations  {for which no efficient algorithm classical computation algorithm is known, and for which it may be that none exist}. Through the preparation, manipulation and processing of quantum information, it should be possible to efficiently solve such problems as quantum simulation \cite{qsim} and the factoring of primes \cite{prime}, as well as allow significant speed-ups for other tasks \cite{grover}. Just as in classical computers, the ability to store quantum information during the computation, to manipulate it while stored and to send it elsewhere when required, will be a critical element of any architecture. Also, such memories must be of very high fidelity. Otherwise, the build up of errors during the computation will yield a useless result. The development of a quantum RAM \cite{qram1,qram2}, and quantum hard drives for long term storage, is therefore a major focus of research into quantum computing.

Because of the many varied and stringent conditions for the storage of information within quantum computers, the use of error correction to create logical quantum memories is required \cite{coderev}. It may, in fact, be advantageous for the logical qubit states to be initialized, processed and read out, all without ever leaving the error correcting code \cite{raus1,raus2,raus3}.

\section{Quantum Error Correction}

So far we have spoken of the possibility of  {combining} many physical memories into a logical memory of higher fidelity. But how may this be achieved? In any quantum memory we typically have access to a system with a large number of degrees of freedom, and we wish to store a relatively small (typically two-level) quantum state. The approach is then to find a two dimensional subspace of the system for which the effects of decoherence is particularly small, and then use this to encode the state we wish to store. By a clever choice of both the system and the subspace, we therefore allow our state to survive a lot longer than if it were encoded in a more exposed manner.

It is exactly this approach that is used with error correcting codes, in both the classical and quantum realm \cite{coderev}. To demonstrate the idea behind them, we start with the simplest example of classical error correction: the repetition code.

Let us suppose that we have a logical  bit that we wish to store, and the physical means to store single bit values, $0$ and $1$. However, this physical memory is noisy. By the time we come to reading out the information, there is a probability $p$ that a bit flip has occurred. The most naive way to store our logical bit is simply to encode it within a single physical memory. However, this method allows the noise to directly affect the logical bit, flipping it with probability $p$ and destroying the integrity of the stored information.

The easiest way to suppress the effects of the noise is simply to copy the logical bit three times. So, if the logical bit in state $0$, we prepare three physical memories in state $000$, and for logical bit $1$ we prepare $111$. At the time of readout, we may find that the three physical bits are in the same state, $000$ for example. The probability that they started in state $0$, and no flips occurred, is $(1-p)^3$. The probability that they started in state $1$, and a flip occurred on each, is $p^3$. Clearly (assuming $p<0.5$), the former is more likely than the latter. We may also find two of the physical bits in the same state and one in the other, $001$ for example. The probability that they all started in state $0$ and suffered a single flip is clearly more likely than the probability that they all started in state $1$ and suffered two flips. This allows us to do a majority voting to correct errors. If the majority of the physical bits are in a certain state, we assume they all started in that state and hence that was the state of the logical bit. The probability that this process fails is related to the smallest error that can turn one majority into the other, i.e. two bit flips. The probability of a logical error is therefore reduced from $p$, to $\sim p^2$. If we used more than three physical bits per logical bit, the error rate would be reduced further. In fact, the logical state can be retrieved with arbitrary fidelity for any $p<0.5$ as the number of physical bits per logical bit is increased.

The repetition code is the  {classical world’s natural} means of storing classical information. It is simply an abstract expression of the fact that classical information is stored in states of many particles, such as the many molecules of ink used to print a letter in a book. Even if the letter is smudged or parts of it are obscured, it has to be perturbed significantly before a passage cannot be read.

The repetition code has a natural extension to qubit encoding only if the quantum state is known. It is only in this case that multiple qubits can be prepared in the same state. However, this is a trivial case, since a known quantum state can be stored easily in a classical computer simply by recording each entry of the density matrix. In any practical application of quantum information, it is the storage of arbitrary quantum states without any prior knowledge that is important. In this case the no-cloning theorem applies \cite{noclone1,noclone2}, preventing such straightforward repetition encoding.

One possible quantum generalization could then be as follows. Say we have an arbitrary qubit state $\ket{\psi} = \alpha \ket{0} + \beta \ket{1}$ (a mixed state would also be valid, but we use pure for simplicity). We can then take two ancilla qubits, each in state $\ket{0}$, and apply a CNOT with each ancilla as target and the original qubit as source. The resulting state is $ \alpha \ket{000} + \beta \ket{111}$, and so $\ket{0}$ is encoded in $\ket{000}$ and $\ket{1}$ in $\ket{111}$. The repetition encoding in the computational basis means that a physical bit flip error rate of $p$ becomes a logical bit flip rate of $\sim p^2$. However, a phase flip acting on any qubit will cause a phase flip to the logical qubit, and no measurable trace of the error will be left. A physical phase flip error rate of $p$ therefore becomes a logical phase flip rate of $\sim 3p$. The encoding protects against bit flips, but allows phase flips to occur more easily. Obviously this does not provide a good quantum code.

A solution to this problem is to concatenate the repetition code in two orthogonal bases. To explain this, we will use the plus/minus basis, the eigenstates of the Pauli $X$ operator on qubits,
\be
\ket{\pm} = \frac{1}{\sqrt{2}}(\ket{0} \pm \ket{1}).
\ee
We will first use three qubits to encode logical plus/minus states $\ket{+}_3 = \ket{+++}$ and $\ket{-}_3 = \ket{---}$ with a standard repetition encoding. The computational basis states in this encoding will take the form,
\bq
\ket{0}_3 &=& \frac{1}{\sqrt{2}}(\ket{+++} + \ket{---}) \\
\ket{1}_3 &=& \frac{1}{\sqrt{2}}(\ket{+++} - \ket{---}).
\eq
This encoding is just that described above, but with the roles of bit and phase flips interchanged. Here it is the phase flips that are suppressed, with a logical rate of $\sim p^2$ and the bit flips whose rate is increased to $\sim 3p$. To suppress the latter we use three groups of three qubits to give a repetition encoding of the computational basis,
\be
\ket{0}_9 = \ket{0}_3\ket{0}_3\ket{0}_3, \,\, \ket{1}_9 = \ket{1}_3\ket{1}_3\ket{1}_3.
\ee
The total encoding is then,
\bq
\ket{0}_9 = \frac{1}{2\sqrt{2}}(\ket{+++} + \ket{---})^{\otimes 3},\\
\ket{1}_9 = \frac{1}{2\sqrt{2}}(\ket{+++} - \ket{---})^{\otimes 3}.
\eq
Both bit and phase flips are suppressed with error rates going from $p \rightarrow p^2$, leading to a lower logical error rate than the physical error rate. Since bit and phase flips are the Pauli operators $X$ and $Z$, respectively (making $Y$ a conjunction of the two) and since all single qubit errors can be decomposed into single qubit Pauli's, this code therefore suppresses all errors. Such suppression in all bases makes this a good quantum error correcting code. It is one of the simplest to achieve such suppression, and is known as the 9-qubit code \cite{9qubit}, for obvious reasons.

From this example, we see that classical error correcting codes do not work for quantum information. However, mashing classical codes together in two orthogonal bases can work. This provides a powerful method of creating quantum codes \cite{css1,css2,coderev}, but are there quantum codes that do not take this form? Can we find a more general framework?

To begin a more general definition of quantum error correcting codes, let us recall the motivation mentioned above: we wish to take a large system, find a subspace for which the effects of decoherence is particularly small, and then use this to encode a logical qubit. To do this we will take a large number of physical quantum memories, whose joint Hilbert space will have the large number of degrees of freedom required. We then choose a subspace of states spread non-locally throughout the physical memories. This is chosen such that, if the state of the system is moved out of the subspace by any small error, the signature of this error can be detected by a cleverly defined set of measurements without disturbing the stored state. These are known as syndrome measurements. Once these have revealed that the error has occurred, it can either be undone or its effects can be accounted for in any subsequent measurements.

An important class of quantum error correcting codes are the stabilizer codes \cite{stabil,coderev}. To define these, one first takes a number of physical memories which each encode $2$-level systems, and then chooses a number of mutually commuting operators acting on these. These operators are known as the stabilizers of the code, they are all products of Pauli operators and so all have eigenvalues $+1$ and $-1$. The stabilizer space of the code is defined to be the mutual $+1$ eigenspace of all stabilizers. It is this subspace that is used to encode the logical state.

Note that the stabilizer operators are Hermitian, and hence are observables. We may therefore consider a measurement of all stabilizers, with each giving a result of $+1$ and $-1$. Since the state of the system should always reside in the stabilizer space, the results of all these measurements should be $+1$. The presence of a stabilizer with outcome $-1$ is therefore a signature that an error has occurred, and so the stabilizer measurements play the role of syndrome measurements. For a well defined stabilizer code, any local error will disturb some stabilizers in this way. An analysis of the results of a measurement of all stabilizers, known as the syndrome, should then be able to undo the effects of the errors by returning the state not only to the stabilizer space, but to its initial state within this space before the error occurred. The resilience of the logical state to decoherence is then much better than the states of the physical memories.

In the next section we will consider a concrete example of a stabilizer code: the planar code. However, it should be noted that there are many approaches to quantum error correction other than the stabilizer formalism. Some are equivalent to stabilizers or generalizations of the approach, and some not. These include CSS codes \cite{css1,css2}, qudit stabilizer codes \cite{qudit1,qudit2,qudit3}, stabilizer codes with weakened conditions \cite{misc}, subsystem codes and topological subsystem codes \cite{sub1,sub2,tsub1,tsub2}, decoherence free subspaces \cite{dfs} and autonomous error correction \cite{kerchoff,nigg,reed,leg}.

\section{Planar code quantum computation}

One particularly interesting form of stabilizer codes are topological stabilizer codes, such as Kitaev's famous toric code \cite{toric,dennis}. For these the physical memories are arranged on a two-dimensional surface, with stabilizers defined on local clusters of these surrounding various points throughout the surface. The eigenstates of the stabilizers are interpreted as occupancies of so-called anyonic quasiparticles, with the eigenvalue $+1$ denoting no anyon is present on the region for which the stabilizer is defined, and $-1$ denoting that an anyon is present. The stabilizer space is therefore the anyonic vacuum. The effect of local errors is to locally create pairs of anyons, and the stabilizer measurement is a measurement of whether anyons exist at each point. Correction is achieved by trying to determine in which pairs the anyons were created (up to an equivalence class) and reannihilating them.

We will now consider the planar code as one particular example of a topological stabilizer code. This will be explained first at an abstract level in terms of the anyonic quasiparticles. At this level, we will see how quantum information can be stored and even how it may be manipulated for universal quantum computation. We will then move to more concrete ground, describing how it may be realized on a spin lattice. This allows us to look at the effects of errors, the means for their suppression and correction and review of the error thresholds that have been found.  {Some readers may prefer to review this concrete section before starting on the abstract.} Finally we will review the idea of using the planar code as a self-correcting quantum memory, by implementing an additional Hamiltonian.

 {It should be noted that there are multiple different ways in which the planar code may be used for quantum computation. Including transversal proposals \cite{dennis} and ones that involve cutting and gluing of codes \cite{clare}. The primary family of proposals, however, are those that use and manipulate `hole'-like defects in the code. These were first introduced by Raussendorf, et al., \cite{raus1,raus2,raus3}, though similar ideas existed previously \cite{pachosabelian}. It is the Raussedorf type approach that we will focus on in this review. However, for reasons of clarity, we will focus on a single-hole encoding of qubits, rather than the original two-hole encoding.}

\subsection{Bosons, fermions and anyons}

\begin{figure}[t]
\begin{center}
{\includegraphics[width=6cm]{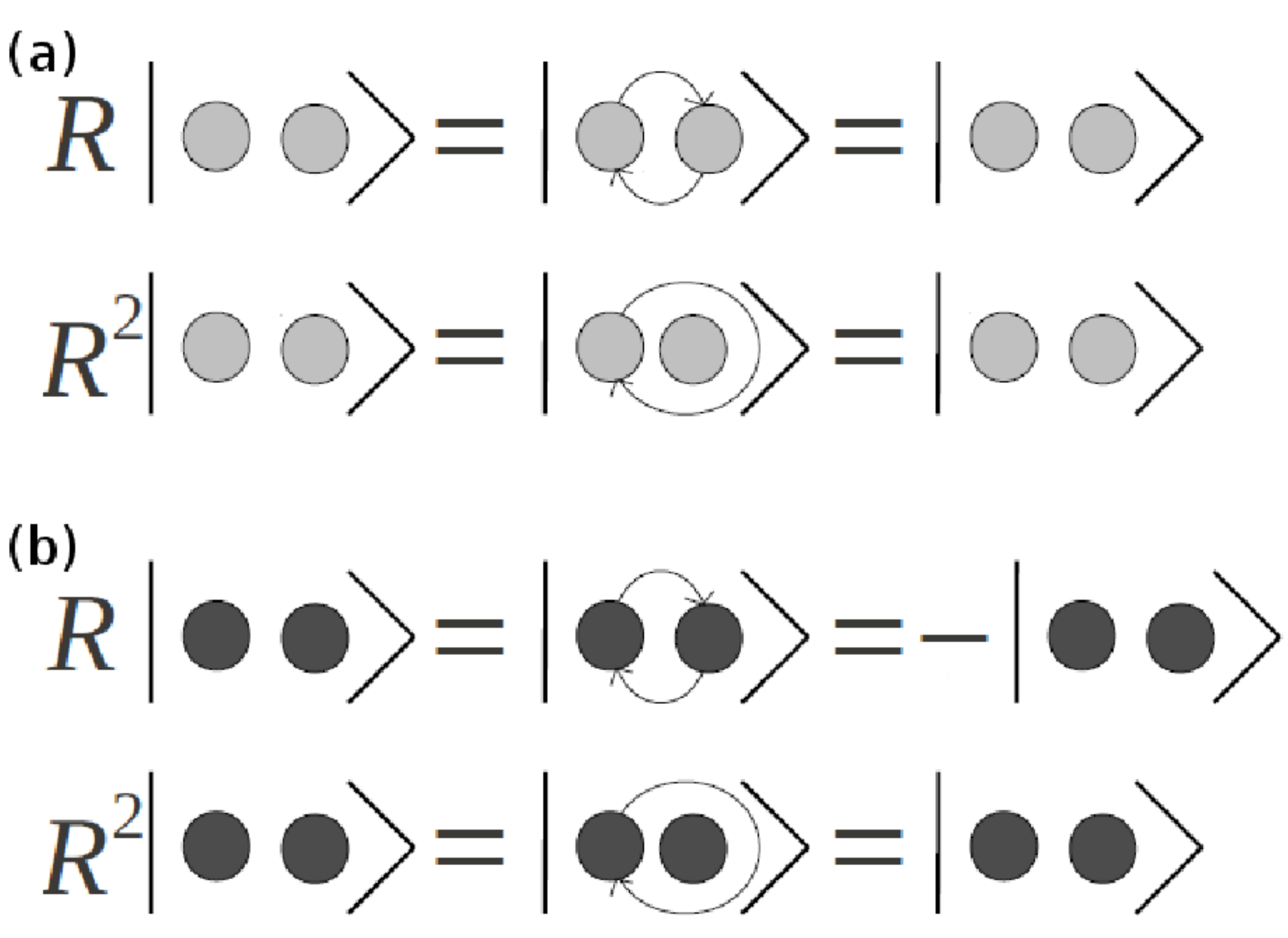}}
\caption{\label{bfbraid} The exchange statistics of (a) bosons and (b) fermions.}
\end{center}
\end{figure}

The properties of bosons and fermions are familiar to any physicist. Their exchange behaviour is summarized in Fig. \ref{bfbraid}. If one has a state of two bosons, and then exchanges them with an exchange operator $R$, the resultant state will be exactly the same as the initial one. This is due to the indistinguishable nature of bosons and the fact that the exchange results a trivial phase (i.e., it results in a phase of $1$). For fermions the situation is the same, except that the exchange leads to a phase of $-1$.

Two successive exchanges of two particles is equivalent to a full monodromy, where one of the particles is braided fully around the other. In both the bosonic and fermionic case the single exchange phase squares to unity. The effects of a full monodromy is therefore trivial in both cases. This is a requirement placed on any point particles in a universe with three spacial dimensions, since a loop around a point in 3D space can be continuously deformed to a loop that is not around a point. A full monodromy is therefore topologically equivalent to a trivial operation, and must have a trivial phase.

In a universe of two spacial dimensions, this restriction no longer holds. A loop around a point can no longer be continuously deformed to a loop that is not around a point, and hence a full monodromy of point particles can have a non-trivial effect. Such particles would be neither bosons or fermions, and so we call them \emph{anyons}.

\subsection{An abstract overview of the planar code}

\begin{figure}[t]
\begin{center}
{\includegraphics[width=6cm]{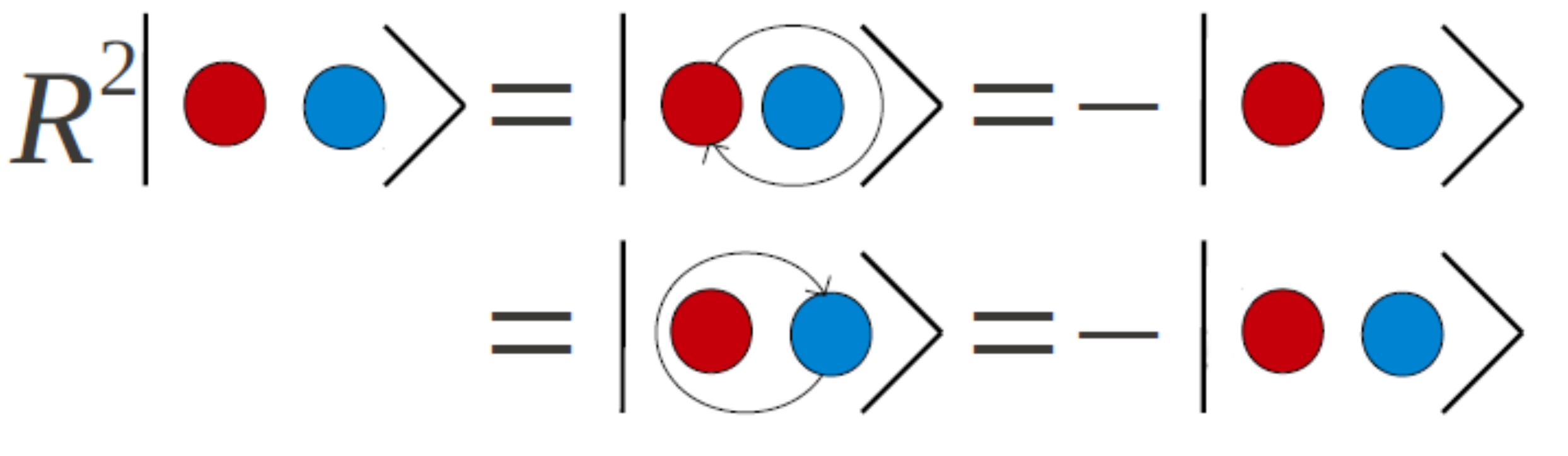}}
\caption{\label{embraid} The exchange statistics of $e$ and $m$ anyons, with $e$ anyons shown in red and $m$ in blue.}
\end{center}
\end{figure}

In the planar code we will imagine that we have access to a 2D `universe' in which there are only two elementary particles, which we call $e$ and $m$. The $e$ is its own antiparticle, and accumulates a trivial phase when exchanged with another $e$. The same is true for the $m$'s. However, their anyonic nature becomes evident when we consider their exchanges with each other. A full monodromy of an $e$ around an $m$ (or equivalently, vice-versa) yields a phase of $-1$, as shown in Fig. \ref{embraid}.

\begin{figure}[t]
\begin{center}
{\includegraphics[width=7cm]{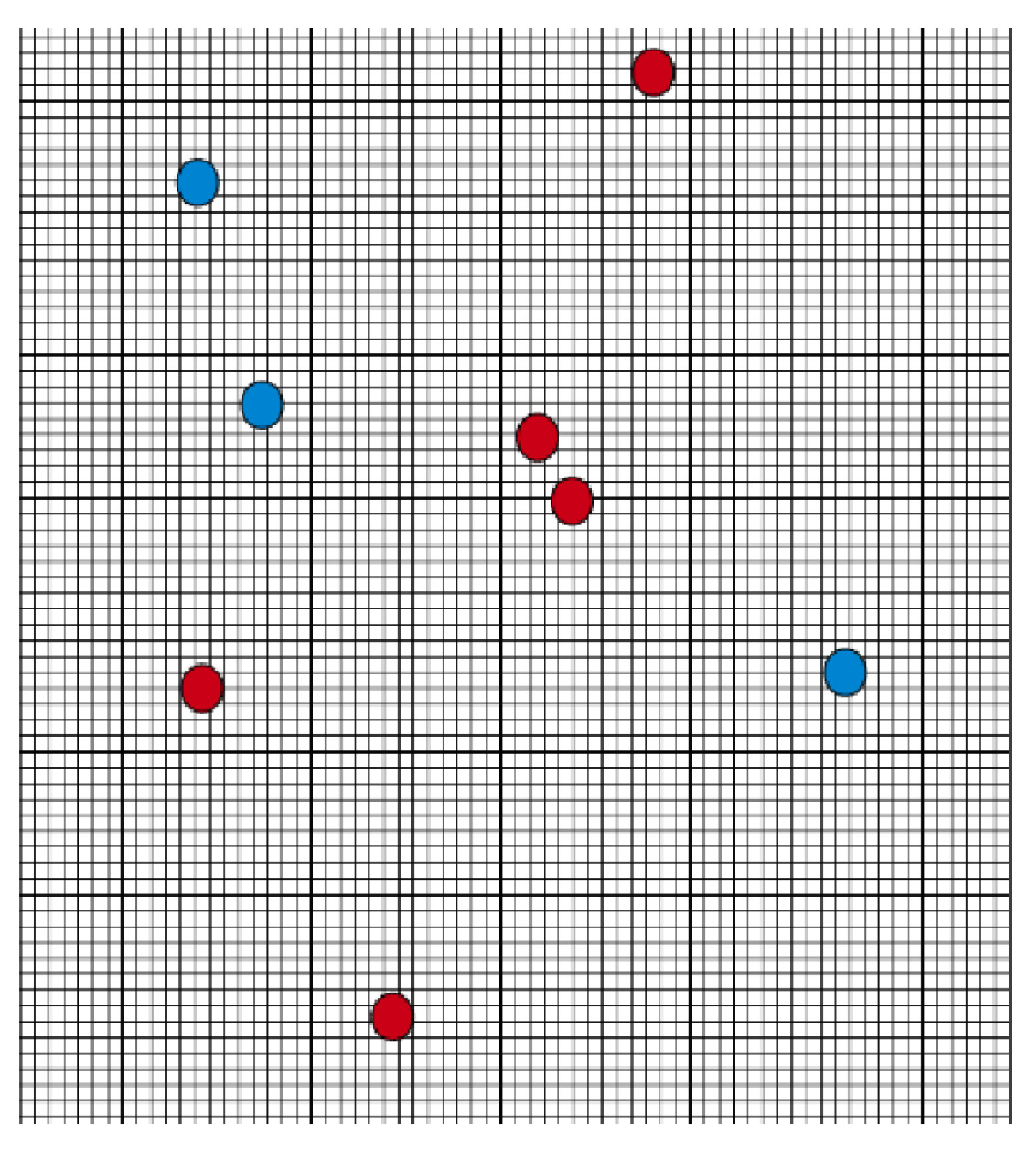}}
\caption{\label{universe}  {The `universe' on which the $e$ and $m$ anyons live, with an example configuration of anyons on the bulk. In general, a configuration of any number of anyons on the bulk is possible. The occupations of the edges will always ensure that the total number of both $e$ and $m$ anyons is even. Note that the lattice background to this figure and the two which follow is not important in itself. It is only to show the distinction between the `universe' and the empty space that surrounds it, and to foreshadow our later discussion in terms of a spin lattice}.}
\end{center}
\end{figure}

We will consider this universe to be square in shape, as depicted in Fig. \ref{universe}. The anyons $e$ and $m$, as point particles, can exist at any localized point in the universe. However, it is also possible for them to exist in a delocalized way on the edges of the universe. Specifically, the top edge of the universe is able to hold a single $e$ anyon. This anyon, if present, cannot be thought of as being at any specific point along the edge. Even if one sees it move from a localized point in the bulk and `fall' of the edge, one cannot think of it as being just  {beyond the edge} from where it disappeared. Rather, the anyonic occupation is spread out over the entire edge. An $e$ anyon can similarly exist on the bottom edge, and an $m$ anyon on each of the two sides. The top and bottom edges appear as hard boundaries to $m$ anyons, as do the sides to $e$'s. An example of a configuration of anyons in this universe can be seen in Fig. \ref{universe}.

 {As we will see in the next section, in reality this universe will be a spin lattice plus a little imagination. We will define operators acting on the spins of the lattice, which will correspond to the stabilizers of an error correcting code. The properties of these operators will allow us to interpret their measurement outcomes as occupations of quasiparticles, with an outcome of $-1$ meaning a particle is present in the region on which the operator is defined, and $+1$ meaning that one is not. The two particle types stem from two different types of stabilizer. By performing operations on the spins of the lattice, we gain certain powers over the anyons of our universe. At this level of abstraction, these powers may seem quite arbitrary. However, when we consider what is really happening in terms of the underlying spin lattice, we will see how they may be justified.} The first power we allow ourselves is the ability to create pairs of anyons from the vacuum. We can create a pair of $e$ or $m$ anyons (but not an $e$ and an $m$ since they are not antiparticles), and also any arbitrary superposition of creating such pairs and creating nothing. We also allow ourselves to move anyons, throw them  {off} their respective edges and annihilate pairs of anyons. We can also measure the anyon occupancy of any point on the bulk as well as the edges, and hence know where  {all the} anyons in the universe are whenever we wish.

\begin{figure}[t]
\begin{center}
{\includegraphics[width=7cm]{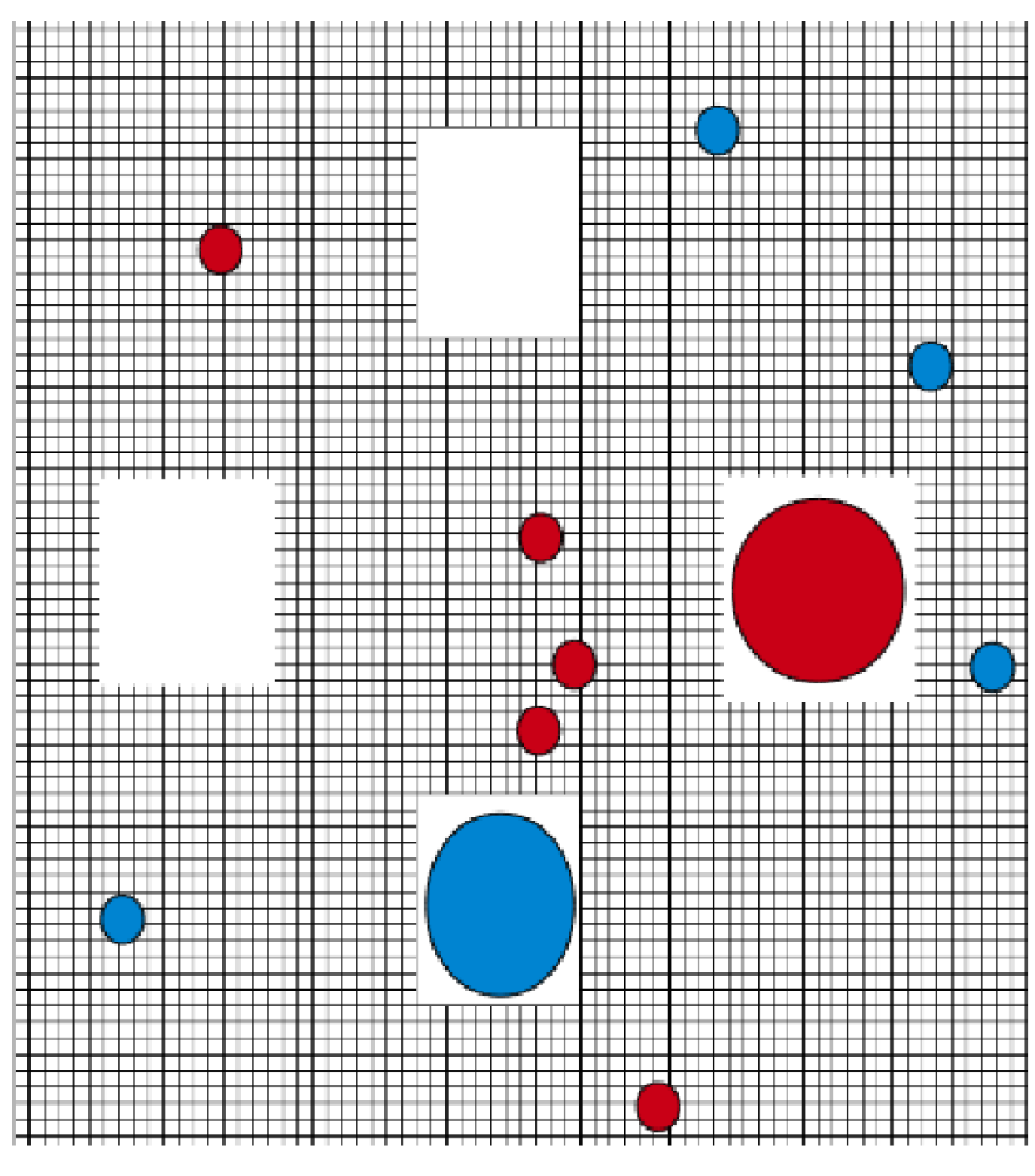}}
\caption{\label{holeuniverse} The universe with four holes, two rough (left and right) and two smooth (top and bottom) and an example configuration of anyons.}
\end{center}
\end{figure}

The final powers we allow ourselves concern `holes' in the universe, as in Fig. \ref{holeuniverse}. There are two types of holes that can exist, which we call rough and smooth,  {due to their appearance when the spin lattice realization of the model is considered}. We allow ourselves to create both types of hole and alter their shape as we wish, allowing us also to move them around. These holes behave in a similar way as edges. Rough (smooth) holes can accommodate a single delocalized $e$ ($m$) anyon, but appear as hard boundaries to $m$ ($e$) anyons. As such, we also allow ourselves to measure the anyonic occupations of the holes.

So, now we have such a universe, is there anything interesting that we can do with it? It turns out that there is: we can use it to perform quantum computation. The first step towards this is to first determine a means of storing the quantum information that we wish to compute. As usual we will do this in the form of qubits: two level systems with states labelled $\ket{0}$ and $\ket{1}$ in the so-called computational basis and $\ket{+}$ and $\ket{-}$ in the orthogonal plus/minus basis. The most basic unitary operations that can be performed on these are the logical $Z$ and $X$ operations, Pauli operators which act on the basis states in the following manner,
\begin{eqnarray*}
Z \ket{0} &=& \ket{0}, \,\, Z \ket{1} = -\ket{1}, \\
X \ket{+} &=& \ket{+}, \,\, X \ket{-} = -\ket{-}, \\
X \ket{0} &=& \ket{1}, \,\, X \ket{1} = \ket{0}, \\
Z \ket{+} &=& \ket{-}, \,\, Z \ket{-} = \ket{+}.
\end{eqnarray*}
We will use holes to store qubits. For each smooth hole we store a single qubit (a smooth qubit), associating the two possible anyon occupations, vacuum and $m$ anyon, with the logical qubit states $\ket{0}$ and $\ket{1}$, respectively. For each rough hole we also store a qubit (a rough qubit), associating vacuum and $e$ anyon occupations with the states $\ket{+}$ and $\ket{-}$, respectively. In order to satisfy the restriction to even numbers of both $e$ and $m$ anyons, any spare anyons are moved  {off} the edge. The bulk is kept anyon free.

The logical $X$ operation maps a qubit in state $\ket{0}$ to one in state $\ket{1}$, and vice-versa. To implement this on a smooth qubit, we therefore need to map a hole containing no anyon to one containing an anyon, and vice-versa. The corresponding operation will obviously be one which creates a pair of $m$ anyons, moves one into the hole, and moves the other  {off} the edge. If the hole previously contained vacuum, it will contain an anyon after this process. If it contained an anyon before, the process annihilates this and results in the vacuum. Similarly, a logical $Z$ can be performed on a rough qubit by creating an $e$ pair, placing one in the hole and moving the other  {off} the edge.

Such operations need to act along all parts of the universe in a path that stretches from the hole to an edge. It is this that gives the qubit its topological protection against errors. No small, local error can effect the qubit as long as the hole is kept far from the edge. Such an error could create a pair of $m$ anyons on the edge of the hole, and cause one to fall in. This would change the anyonic occupation of the hole, and hence appear to have caused an $X$ error. However, the presence of the other anyon near the hole is a clear signature of what happened. Any measurement of the anyon configuration would reveal that the error occurred, and the spare anyon may be thrown in the hole to annihilate its partner and undo the problem.

Logical $Z$ operations for rough qubits and logical $X$ operations for smooth qubits require a phase of $-1$ to be generated if an anyon resides in the associated hole, and no phase to be generated otherwise. To achieve this, we can use the braiding properties of the anyons. To apply a $Z$ to a smooth qubit we first create a pair of $e$ anyons, braid one around the hole, and then annihilate the pair. If the hole contained vacuum (and hence the qubit was in state $\ket{0}$, this process has trivial effect. If it contained an $m$ anyon (qubit in state $\ket{1}$), the braiding yields the required phase of $-1$. Similarly an $X$ can be applied to rough qubits by creating an $m$ pair, braiding one and annihilating.

\begin{figure}[t]
\begin{center}
{\includegraphics[width=5cm]{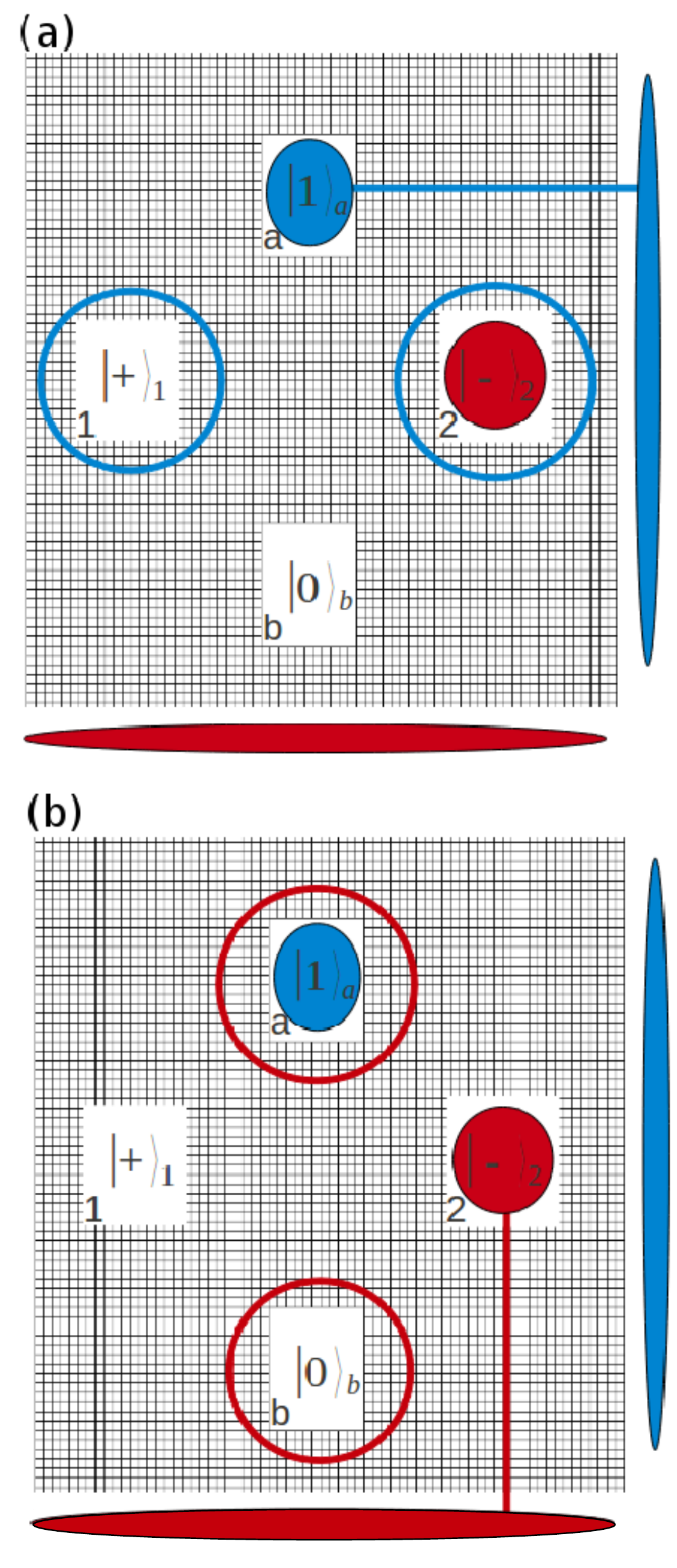}}
\caption{\label{encode} Four holes in which four qubits (two rough and two small) are stored. (a) The paths along which $m$ anyons must be moved in order to perform logical $X$ operations. For smooth qubits these stretch from hole to edge, changing the occupancy of both. For rough qubits these encircle the hole. (b) Similarly, the paths required for $e$ anyons to perform logical $Z$ operations.}
\end{center}
\end{figure}

Such operations also need to act along a path, this time one which completely encircles a hole. Making the holes large therefore topologically protects the qubits. Any local error can only create and move anyons around a small portion of the hole. The probability of many such errors conspiring to braid an anyon all (or most of) the way around is suppressed with the hole's diameter. The encoding and the logical operators are depicted in Fig. \ref{encode}.

Since any logical error can be decomposed into $Z$ and $X$ errors, the fact that both of these are topologically protected for both qubit types (rough and smooth) is sufficient to show that the qubits are completely topologically protected so long as the holes are large and well separated \cite{raus1,raus2,raus3,fowlerrev,daverev,fowlerrev2}. However, a full treatment of errors is deferred until later.

It should also be noted that the logical encoding allows fault-tolerant preparation of smooth qubits in the $\ket{0}$ and $\ket{1}$ basis states, and rough qubits in the $\ket{+}$ and $\ket{-}$ basis states. This is done by first clearing all anyons away before defining the holes. It is then known that the holes must contain the vacuum, and any local errors that may alter this will leave the same signature as described above. This gives the preparation of smooth $\ket{0}$ states and rough $\ket{+}$ states, and allows preparation of smooth $\ket{1}$ states and rough $\ket{-}$ states by application of an $X$ and $Z$, respectively.

An entangling operation can be performed on the stored qubits in a very straightforward manner. Since we are allowed to move the holes, let us consider braiding a smooth hole in a full monodromy around a rough hole (or vice-versa). The effects of this obviously depend on the anyonic occupations of the two holes. If the smooth hole contains the vacuum (and hence is in logical state $\ket{0}$), the braiding will be trivial. If it contains an $m$ anyon (and hence is in logical state $\ket{1}$), the effect is to braid an $m$ around the rough hole, generating a phase of $-1$ iff that hole contains an $e$ anyon (and hence is in state $\ket{-}$). This corresponds to a logical $X$ operation on the rough qubit. The total effect of the braiding of the two holes is therefore to implement the entangling CNOT gate, with the smooth qubit as the control and the rough as the target. Like the logical $X$ and $Z$ operators, the CNOT is also topologically protected. Local errors cannot cause the braiding of the hole to take place in error if the holes are sufficiently large. Also, though the operations used to move the holes may be imperfect and cause local creation and movement of anyons, these can be identified and corrected as long as the movement is sufficiently slow.

The ability to realize a fault-tolerant entangling gate demonstrates the power that this model has to process quantum information. However, the gate set that we can realize through the creation and braiding of the anyons is still not universal. For that we require something more.

To see what we can use to achieve universality, consider in general some non-universal gate set, such as the Clifford group of operators. Add to this set another operator, namely preparation of a given single qubit state. If this new set of operators is universal, the state can be called a `magic state' \cite{magic1,magic2}. The ability to prepare this state and use it as an ancilla boosts a non-universal gate set to universal.

Given the gates that we can already achieve, it can be shown that the required magic states for smooth qubits are of the form,
\be
\ket\theta = \frac{1}{\sqrt 2} ( \ket{0} + \exp{i \theta} \ket{1} ),
\ee
for $\theta = 0, \, \pi/4, \, \pi/2, \, \pi$. With similar states required for rough qubits, except that $\ket{0}$ is changed for $\ket{+}$ and $\ket{1}$ for $\ket{-}$. To prepare these, smooth qubits can be moved near to side edges, shortening the distance required to perform the logical $X$ operator. In fact, using the spin lattice realization of the next section, it will allow the logical $X$ basis to be accessed through operations on a single spin \cite{fowlerrev,wootton1,wootton2,wootton3}. The manipulations required to prepare a state such as the above may then be achieved through $X$-basis measurements. Some of these may lead to the creation of unwanted $e$ particles, but these will be removed through error correction.

Of course, making something easy for us to do means making it easy for nature also. If we can perform operations in the $X$ basis, so too can nature perform errors. However, note that the states with $\theta=0, \, \pi$ are eigenstates of the $X$ operator, and hence are immune to $X$ errors. We needn't worry about these as long as the holes are large enough to prevent $Z$ errors. The same does not apply to the $\theta=\pi/4, \, \pi/2$ states. The errors can be minimized by moving the holes away from the edge as soon as the preparation is finished, but a significant amount will still occur. To deal with these, it is found that distillation can be used \cite{magic1,magic2,raus1,raus2,raus3}. This takes many noisy copies of the states and uses the original (and non-universal) gates available to the model to distil these down to a single copy with arbitrary accuracy.

The magic states can therefore be prepared fault-tolerantly. Added to the other fault-tolerant gates possible through the manipulation of the anyons, this allows the model to achieve the holy grail of fault-tolerant universal quantum computation.

\subsection{Realizing the planar code on a spin lattice}

In order for the ideas developed above to bear fruit, we must find some way to simulate this anyonic universe in the lab. To do this, we need a lattice of two-level quantum systems. It is usual to call these systems `spins', and hence we use a `spin lattice'. However, the systems need not actually have any spin associated with them. Any two-level quantum systems will do, so long as they can be manipulated in the required way. As such, it is in fact photon based systems that are responsible for most experimental progress in the planar code so far \cite{ex1,ex2,ex3}. 

\begin{figure}[t]
\begin{center}
{\includegraphics[width=7cm]{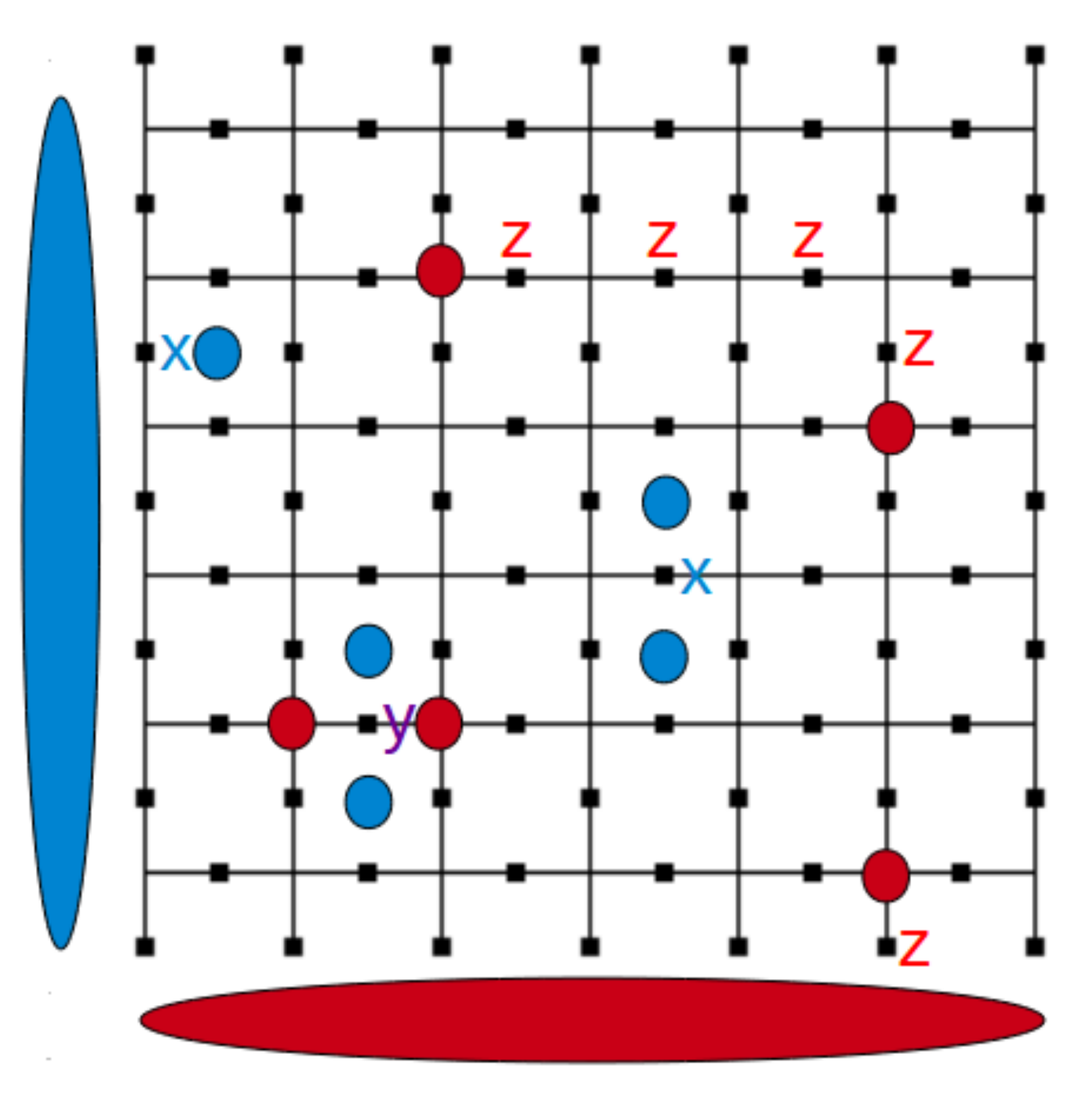}}
\caption{\label{lattice} The lattice on which the planar code is defined. A selection of single spin errors and their effects in creating and moving anyons are also shown.  {Note that the edge anyons denote anyons that have been moved off the edge by errors, rather than the edge occupations themselves. }
}
\end{center}
\end{figure}

It is here that the physical quantum memories enter the discussion, since it is these that will serve as the spins. To implement the planar code we take many physical quantum memories, each of which have a noise rate greater than the required level, and place them in a lattice such as that of Fig. \ref{lattice}. The size of this lattice is characterized by the  number of plaquettes along the horizontal direction or the number of vertices along the vertical. These should be of the same order. We will take them to be equal, and denote them by $L$.

Multi-spin observables are defined on the spins, and correspond to occupations of $e$ and $m$ on different points on the lattice. The steps developed in the previous section can then be used to create a number of logical quantum memories, in which qubits can be stored at a greatly reduced error rate and used for fault-tolerant quantum computation.

The two bases we will use when studying the states of the spins are the computational basis and plus/minus basis, as used for the logical qubits above. The Pauli operators for the spins will be denoted $\sigma^x$, $\sigma^y$ and $\sigma^z$.

The observables used to define the occupancies of $m$ anyons are associated with each plaquette (or face) of the lattice. Specifically, consider the spins of the lattice in the computational basis. Each plaquette is surrounded by four spins (three for those near the edge). If, for a given plaquette, an even number of these spins are in the state $\ket{1}$, we say that that plaquette does not hold an anyon. If there are an odd number of $\ket{1}$'s, we say that it holds an $m$ anyon.

Similarly, the occupancies of $e$ anyons are defined on the vertices of the lattice, which are also surrounded by four spins, or three near the edge. If, for a given vertex, an even number of these four spins are in the state $\ket{-}$, we say that that vertex does not hold an anyon. If there are an odd number of $\ket{-}$'s, we say that it holds an $e$ anyon.

The occupations of the edges are defined in the same way. If an even number of the spins on the left (right) edge are the state $\ket{1}$, the left (right) edge holds no anyons. If an odd number, it holds an $m$ anyon. An even number of $\ket{-}$'s on the top (bottom) edge means no anyon and an odd number means an $e$ anyon is present.

The definition of the anyons is no more complex or mysterious than this. The $e$ and $m$ anyons simply correspond to the parity of $\ket{-}$'s and $\ket{1}$'s around vertices and plaquettes, respectively. From these simple definitions, we find that the anyons will have all the properties described in the previous section. However, before we continue, let us state the definitions in a more formal way. To do that, the following observables are defined for the spins around each plaquette, $p$, and vertex, $s$,
\be
A_s = \prod_{i \in s} \sigma^x_i, \,\,\, B_p = \prod_{i \in p} \sigma^z_i,
\ee
where $i \in s$ denotes the spins $i$ that surround the vertex $s$, etc. These are the stabilizer operators of the code. Each observable $B_p$ has eigenvalues $+1$ and $-1$. The $+1$ ($-1$) eigenspace is formed by states for which there is an even (odd) number of $\ket{1}$'s around the plaquette $p$. The $+1$ and $-1$ eigenspaces therefore correspond to states of vacuum and an $m$ anyon on the plaquette $p$, respectively. These eigenvalues for the $A_s$ similarly correspond to the vacuum and an $e$ anyon on the vertex $s$, respectively.

Note that these operators all mutually commute. This allows the configuration of $e$ and $m$ anyons to both be well defined at the same time. Measurement of $e$ occupations on vertices will not disturb $m$ occupations on plaquettes, or vice-versa. The stabilizer space of the code is defined as the space of states that satisfy the conditions,
\be
A_s \ket{\psi} = \ket{\psi}, \,\,\, B_p \ket{\psi} = \ket{\psi}.
\ee
The stabilizer space is therefore the mutual $+1$ eigenspace of all the plaquette and vertex operators, and corresponds to the anyonic vacuum.

In addition to the stabilizer operators are the edge operators, which act on the stabilizer space. The left and right edge operators are products of $\sigma^z$'s down the left and right edges of the code, respectively. The top and bottom operators are products of $\sigma^x$'s along the top and bottom. These commute with all stabilizers, and the product of left and right commutes with the product of top and bottom to ensure that the net anyon occupation  {off} the entire edge is well defined, however they do not mutually commute. This will be important for the edge encoding considered later.

Let us consider the system in a state within the stabilizer space, and look at the effects of local errors. Since all operators can be decomposed into the Pauli basis, it is sufficient to consider only Pauli operators. A $\sigma^x$ applied to a spin $i$ will anticommute with two $B_p$ operators, those for the plaquettes that neighbour the spin. The effect is therefore to flip the sign of these two stabilizers, creating an $m$ anyon on the two plaquettes. The operator commutes with all other stabilizers, and so has no further effect. The application of a $\sigma^z$ on any spin will similarly create $e$ anyons on neighbouring vertices, and a $\sigma^y$ will create both an $e$ and an $m$ pair.

Now let us take a state that is not within the stabilizer space, but already has an $m$ anyon present on the plaquette $p$. What is the effect of a $\sigma^x$ applied to the spin shared by the plaquette $p$ and its neighbour $p'$? It will flip the sign of both, returning the state to the $+1$ eigenspace of $B_p$, but placing it in the $-1$ eigenspace of $B_{p'}$. This $\sigma^x$ therefore moves the $m$ anyon, and $\sigma^z$'s will similarly move $e$ anyons.

In general, the effects of Pauli operators are to create, move and annihilate anyons, as shown in Fig. \ref{lattice}. Note that to create a pair of $m$ anyons, move them and then annihilate them, a loop of $\sigma_x$ operations is required. This loop will be equal to a product of the $A_s$ stabilizers for the vertices enclosed by a loop. The same is true for $e$ anyons, except that the loop will be one of $B_p$ stabilizers. Since the stabilizer space is an eigenstate of all stabilizers, such closed anyon loops will have trivial effect.

 {The anyonic braiding of the $e$ and $m$ anyons is due entirely to the anticommutation of $\sigma_z$ and $\sigma_x$ operations. For example, consider the creation and movement of an $e$ pair by the application of a string of $\sigma_z$'s. Creating a $m$ pair, braiding one of them around of the $e$'s, and then reannihilating the $m$ pair, can be achieved by a loop of $\sigma_x$ operators that is equivalent to a product of $A_s$ stabilizers. The fact of the braiding means that the loop will enclose the $e$ that was braided around, and so the loop of $\sigma_x$'s will overlap with the string of $\sigma_z$'s on an odd number of spins. The resulting anticommutation results in the phase of $-1$ acquired through the braiding.}

To store qubits in the way described in our earlier abstraction we need to make holes in the code. If the holes were to be static, this would be done simply by modifying the lattice on which the code is defined. Instead of the standard square lattice, a square lattices with holes could be used. However, since we wish holes to be a resource that can be created and moved at will, we will need to specify a means by which this can be done.

Let us begin with the standard square lattice planar code. Creating a hole means choosing an area of the lattice on which we no longer wish to enforce stabilizers. This will mean that the spins within that region will not be part of the code for as long as the hole resides in that place. As such, they must be disentangled from the rest of the system.

To create a smooth hole, we first choose a selection of plaquette stabilizers that we no longer wish to enforce, and then determine which spins are acted upon only by those plaquettes and no others. These are then disentangled by performing measurements in the $\sigma^x$ basis, and hence the plus/minus basis, of the spins. To keep things tidy, all spins are rotated (if needed) to the $\ket{+}$ state after this measurement.

\begin{figure}[t]
\begin{center}
\includegraphics[width=8.5cm]{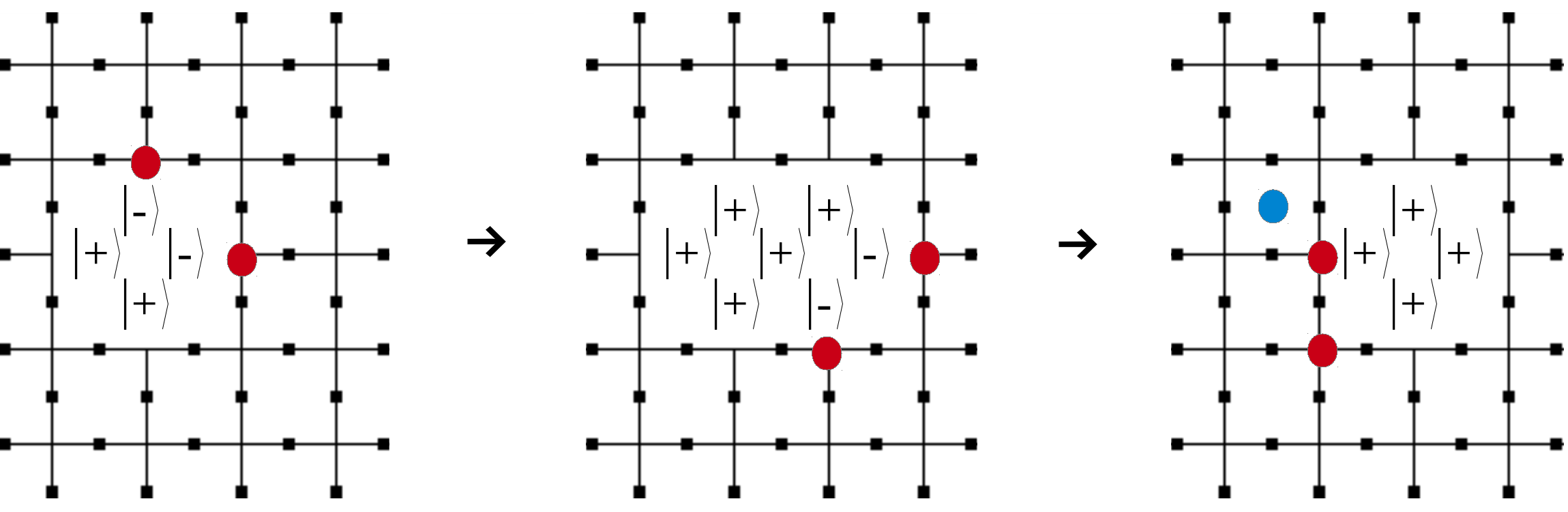}
\caption{\label{holes1} The creation and movement of a smooth hole is shown, first by creation, then expansion and finally contraction. Creation is performed by measuring the spins within the hole in the $\sigma^x$ basis. The parity of the number of spins with outcome $\ket{-}$ will be equal to the parity of the number of $e$'s initially within the vertices contained by the hole, and hence will be even if the state was initially in the stabilizer space. For every spin with outcome $\ket{-}$, the corresponding truncated stabilizer will hold an $e$ anyon. These must be removed, or otherwise taken into account, and the spins in state $\ket{-}$ should be rotated to $\ket{+}$. Expansion is simply achieved through more creation. The additional $e$ anyons will only arise on the expanded boundary. To contract a hole, previously unenforced stabilizers need to be measured. If these contain $m$ anyons, they must be placed in the hole. If they contain $e$ anyons, these must be annihilated. Creation and movement of rough holes are achieved similarly, with the roles of plaquettes and vertices, and of $e$ and $m$ anyons, reversed. Also, the $\sigma^z$ basis is used in place of the $\sigma^x$.
}
\end{center}
\end{figure}

There will be, in general, some vertex stabilizers with support only on the disentangled spins. These will be of vertices that reside within the hole. These stabilizers are also no longer enforced. There will also be stabilizers with partial support on the disentangled spins. Specifically, those residing on all points of the perimeter of the hole, but not on the corners, will act on one disentangled spin. The definition of these stabilizers is then changed accordingly, from four-body operators to three body operators acting only on the spins in the code. If the original stabilizer was in its $+1$ eigenspace, and hence the vertex held no anyon, the same may not be true for the reduced stabilizer. The creation of the smooth hole will therefore result in spare $e$ anyons that need to be cleared up. The end result is then to create a large plaquette in the code, which is the composite of the smaller plaquettes whose stabilizers are no longer enforced. Like any other plaquette, this one can either hold vacuum or an $m$ anyon. However, unlike the others, it has no stabilizer associated with it. As such, the dimension of the stabilizer space is increased by the presence of the hole, since it holds states associated with both possible occupations. The creation of a smooth hole is depicted in Fig. \ref{holes1}.

If the code was initially in the stabilizer space, with anyonic vacuum everywhere, the hole will be initialized in the vacuum state. There will always be an even number of spare $e$ anyons created, which can be annihilated with each other without affecting the hole state. In general, the hole will hold vacuum if the component plaquettes initially held net vacuum (an even number of $m$'s) and will hold an $m$ if the component plaquettes initially held a net $m$.

The creation of a rough hole is similar. A selection of vertices are chosen for which the stabilizers will no longer be enforced. The spins touched only by these are identified and measured in the computational basis of the spins. Stabilizers for plaquettes within the holes are no longer enforced, and those bordering it are reduced if required. This will result in an even number of spare $m$ anyons to be cleared up. The hole is then effectively a large vertex (though it is not often depicted as such for sake of clarity) that is the composite of those vertices that are no longer enforced. It can hold either the vacuum or an $e$ anyon, with its initial state dependent on the previous occupations of the component vertices.

Since the presence of a hole increases the stabilizer space, they can be used to store quantum information. Specifically, each hole can store a logical qubit. The $\ket{0}$ and $\ket{1}$ states of a smooth hole are defined by the vacuum and $m$ occupancies, respectively. The logical $Z$ operator is therefore the product of the (now no longer enforced) stabilizers of the component plaquettes, and takes the form of a product of $\sigma_z$ operators that encircle the spin. Similarly the logical $X$ for rough holes, where the $\ket{+}$ and $\ket{-}$ states are defined by the presence or lack of an $e$ anyon, is the product of stabilizers for the component vertices.

Note that it was earlier stated that products of stabilizers act trivially on the stabilizer space, and yet here we find that such a product is a logical operator. The resolution to this apparent contradiction is that the stabilizers for the component plaquettes and vertices are no longer a part of the code. It is only products of stabilizers within the code that act trivially. Anyon loops are therefore not trivial if the anyons have braided around holes, but are trivial otherwise.

\begin{figure}[t]
\begin{center}
{\includegraphics[width=8cm]{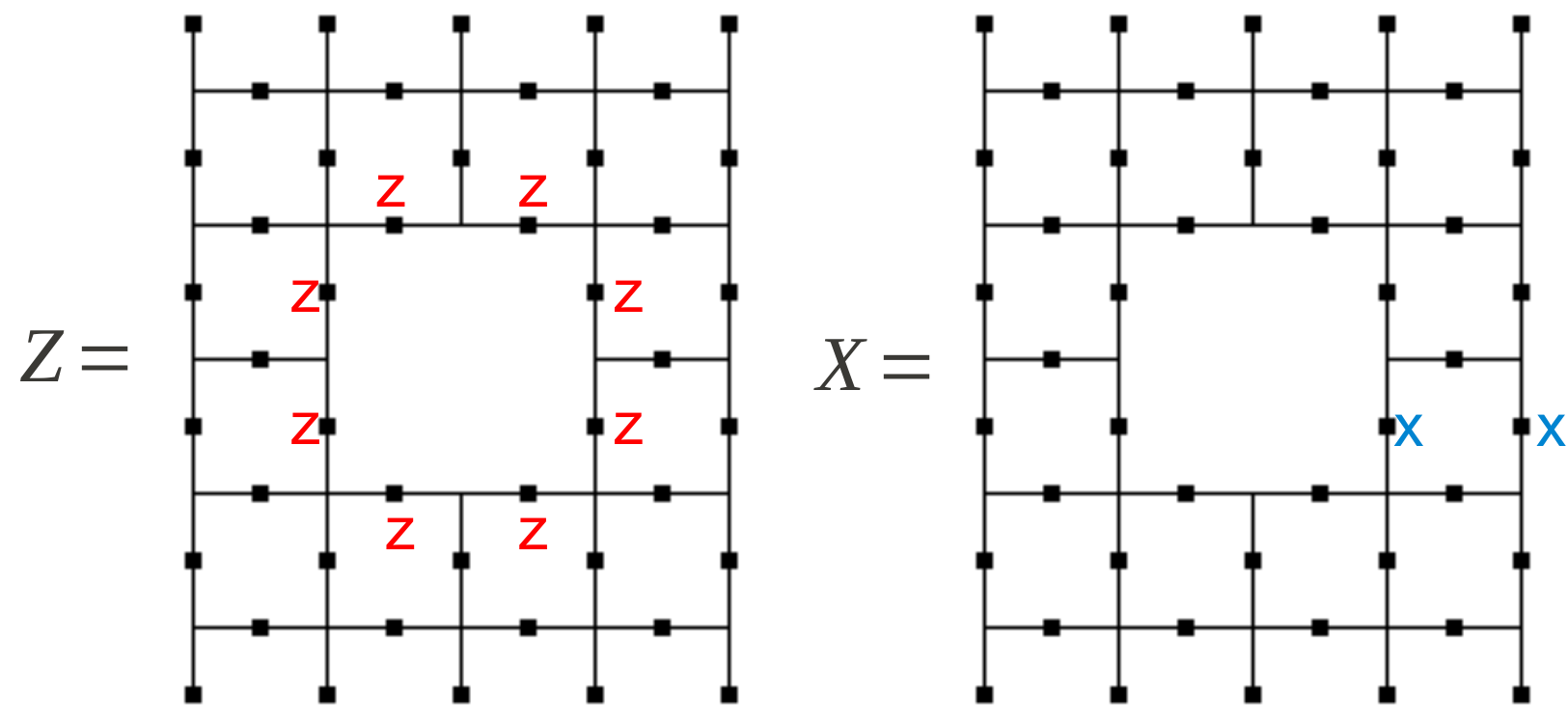}}
\caption{\label{holes2} Logical operators for smooth holes. The $Z$ operator is a loop of $\sigma^z$'s around the hole, and the $X$ is a string of $\sigma^x$'s from the hole to the edge. For rough holes the roles of $Z$ and $X$, and $\sigma^z$ and $\sigma^x$, are interchanged.
}
\end{center}
\end{figure}

The logical $X$ operator for smooth holes requires the occupancy of the hole to be changed. It does this by creating a pair of $m$ anyons, and placing one within the hole. However, the requirement for logical states to lie within the stabilizer space means that the other $m$ anyon cannot reside on any plaquette for which the stabilizer is still enforced. The logical operator must therefore take this spare anyon to, and place it over, the edge of the code. The logical $X$ is therefore a string of $\sigma^x$ operators from a smooth hole to the left or right edge. The logical $Z$ for rough holes is similarly a string of $\sigma^x$'s from the rough hole to the top or bottom edge. These logical operators are depicted in Fig. \ref{holes2}.

The movement of a hole is achieved through its expansion and contraction. To expand a smooth hole, simply choose more plaquette stabilizers on the boundary of the existing hole that you wish to no longer enforce. Identify the spins that are no part of no enforced plaquette stabilizer, and measure them in the $\sigma^z$ basis. Finally redefine the reduced vertex stabilizers on the edge of the hole and measure them, finding the spare $e$ anyons that have arisen on these and removing them. The movement of a smooth hole is depicted in Fig. \ref{holes1}

For this final step, there is an additional complication than must be considered. When a hole is being moved, its logical qubit will typically be in an arbitrary state. As such, the application of unwanted logical $Z$ errors must be avoided. However, consider the case that the hole is expanded in all directions, and so retains none of its initial boundary. Spare $e$ anyons will be created randomly around the entire perimeter of the hole. Two completely valid ways to annihilate these with each other will be possible. However, the operators required to implement these differ by a factor of the $Z$ operator. As such, we are equally likely to annihilate the anyons the right way, which does not implement a $Z$ on the stored qubit, and the wrong way which does.

On the other hand, consider an expansion that only pushes back part of the hole's perimeter. In this case it is known that the operator for annihilation of the spare anyons should only have support on that part of the perimeter that was expanded.  The choice of how to annihilate is then unique (up to trivial factors of stabilizers) and will not implement a logical $Z$. As such, it must be remembered that holes are to be expanded slowly. A small portion must be expanded, then error correction must be run to annihilate any anyons that have arisen during the expansion, either because of the expansion or because of spin errors, then the hole can be expanded a little more. All the same arguments also apply to rough holes.

\subsubsection{Edge encoding}

The hole based encoding is required to perform the quantum computation scheme described in the earlier abstraction. However, there is a simpler encoding that can be used if the planar code is required only to serve as a quantum memory. It's simplicity also means that it is the basis of proof-of-principle experiments.

\begin{figure}[t]
\begin{center}
{\includegraphics[width=7cm]{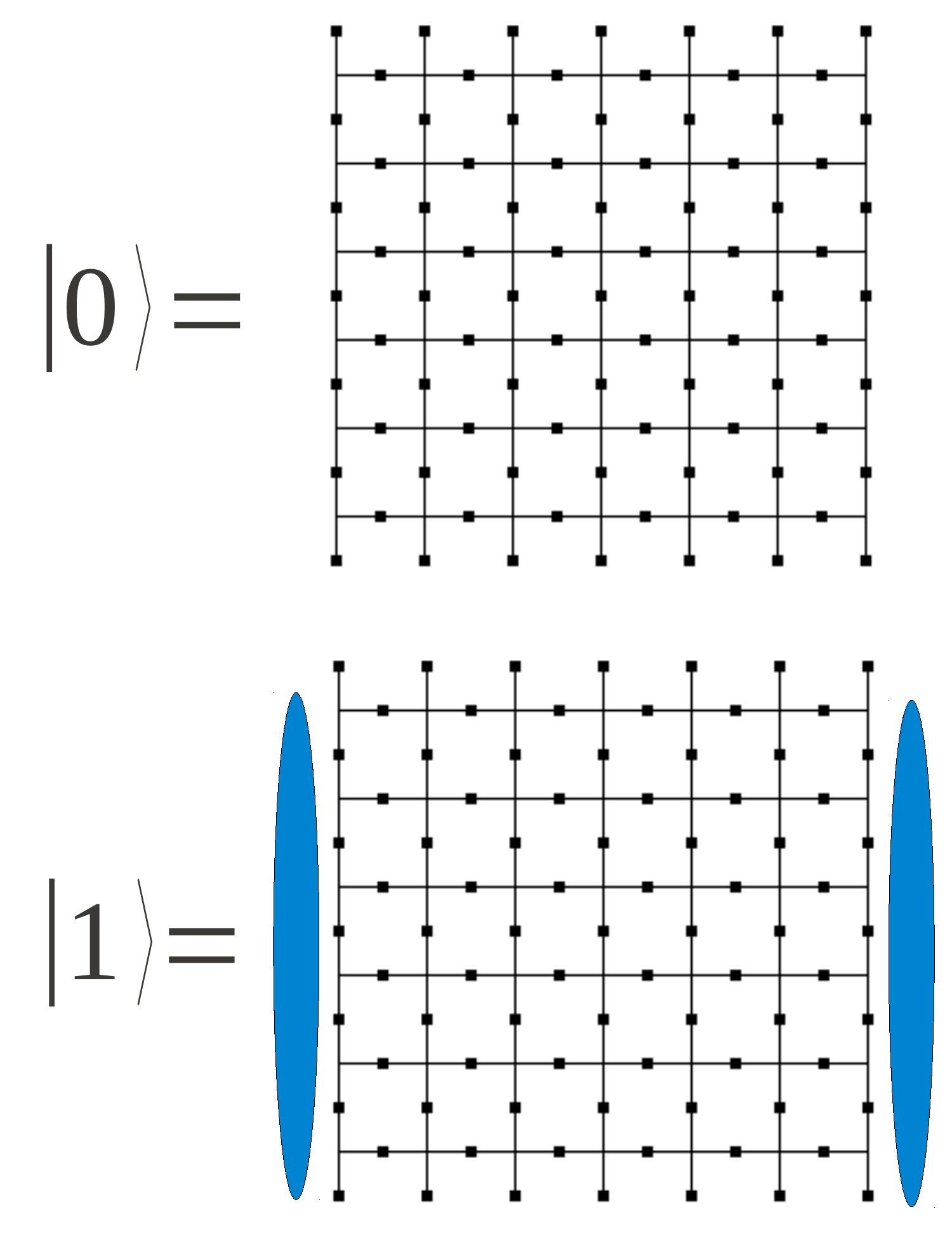}}
\caption{\label{edge}  {A depiction of the computational basis encoding using the edges. The state $\ket{0}$ is defined as a state with no $m$ anyons on the left and right edges, and $\ket{1}$ as a state with $m$ anyons on these edges. The top and bottom edges are correspondingly a superposition of both holding vacuum and both holding $e$.}}
\end{center}
\end{figure}

The simplest way to store qubits in this code is using the edge occupations. A logical $\ket{0}$ can be stored in the state with vacuum on the left and right edges, and $\ket{1}$ in the state with an $m$ anyon on these edges. The logical $\ket{+}$ ($\ket{-}$) is stored using vacuum ($e$ anyons) on the top and bottom edges. This is depicted in Fig. \ref{edge}. Because the occupation operator for the top or bottom edges does not commute with that for the left or right, these are not independent qubits as they would be for a hole based encoding. Instead they are the logical $Z$ and $X$ basis states of a single qubit. The logical $X$ and $Z$ operators correspond to the edge occupation operators and so stretch across the entire code. In order to perform a logical error, nature must therefore create a pair of anyons and transport them the width of the code, with each going to opposite edges. As for the hole based encoding, these means that errors are suppressed for large system size.

Though the hole based encoding is the primary approach for quantum computation in the planar code, other approaches have been proposed using this edge encoding. These include transversal application of gates \cite{dennis}, and so-called `lattice surgery', which interacts the logical qubits stored on different codes by combining the codes and then splitting them \cite{clare}.

\subsubsection{Stabilizer measurements}

By now it should be obvious that the workhorse of the planar code, when used solely as a quantum memory or when quantum computation is performed, is measurement. It is by measurement that errors are detected such that they may be corrected, and it is measurement that makes and moves the holes. All other operations need only act on single spins. It is the measurements that drive their effects non-locally and allow them to have computational power.

Most important of the measurements are the stabilizer measurements, which are the only multi-spin aspect of the proposal. Since there are four-spin observables, their measurement is obviously not straightforward. The naive approach would be to deduce the outcome of the four-spin measurement by making four single-spin measurements. Measurement of $B_p$, for example, determines whether there are an odd or even number of $\ket{1}$'s around the plaquette $p$. This could be deduced by measuring each of these spins in the computational basis and counting the number of $\ket{1}$'s. However, this approach would not commute with the neighbouring $A_s$ operators, and so would cause $\sigma^z$ errors. Measuring all plaquette stabilizers in this way would then leads to logical $Z$ errors, as would measuring all vertex stabilizers like this lead to logical $X$ errors.


A means to measure the stabilizers must be therefore be used that is both realistic and commutes with all other stabilizers. For the most part, direct measurement of the four-spin operator is not realistic, though some exceptions exist \cite{fabio,solgun}. The generally favoured approach is therefore to use ancilla spins located at the centre of every vertex and plaquette. Code spins then couple to these using CNOT gates in such a way that the outcome of the four-spin measurement is mapped onto the single spin ancilla state. Single spin measurement may then be used to reveal its value.

To be more specific, an anyon is located on a plaquette if the state is in the $-1$ eigenspace of $B_p$, and so if there are an odd number of spins around the plaquette in state $\ket{1}$. To determine this, we take a spin initialized in state $\ket{0}$ and place it inside the plaquette. CNOT's may then be performed with each spin around the plaquette as the source and the ancilla as the target. A CNOT performs a $\sigma^x$ on the target only if the source is in state $\ket{1}$, and so as many $\sigma^x$'s will be performed on the ancilla as there are $\ket{1}$'s around the plaquette. An even number will cancel each other out, leaving the ancilla in state $\ket{0}$. An odd number acts as a single $\sigma^x$, leaving the ancilla in state $\ket{1}$. Measuring the ancilla and obtaining the result $\ket{0}$ or $\ket{1}$ then means that the plaquette holds the vacuum or an anyon, respectively.

For vertices, it is the parity of $\ket{-}$'s that determines whether an anyon is present. As such, a phase controlled NOT is required, which performs a $\sigma^x$ on the target only if the source is in state $\ket{-}$.  This is obtained from a CNOT simply by performing a Hadamard on the source spin before and after the CNOT's application. The vertex occupation is read-out similarly to the above by placing a $\ket{0}$ initialized ancilla in the center of the vertex, and performing phase controlled NOT's with each vertex spin as the source and the ancilla of the target. The final ancilla state then represents whether there was vacuum in the vertex, or an anyon. Note that the CNOT's and phase controlled NOT's do not individually commute with the surrounding stabilizers. However, once the four operations required for any plaquette or vertex are applied, the composite operation does commute with all stabilizers. This form of measurement therefore does not disturb other plaquettes and vertices, or cause logical errors (at least, when done perfectly).

With this in mind, we can reduce the entire planar code down to a few simple ingredients. We need to place a large number of physical memories in a two-dimensional lattice. We need to be able to act on these with single spin operations (measurements and rotations in the $\sigma^x$ and $\sigma^z$ bases, and the Hadamard). And we need to be able to perform CNOT's between neighbouring physical memories. If these can be performed experimentally, the code can be implemented. This will allow us to achieve either a straightforward quantum memory, or quantum computation. This is subject of course to the noise level, as we will now discuss.

\subsection{Errors and thresholds}

When considering quantum memories, one important figure of merit to be considered is the lifetime. This will depend on various parameters, such as the temperature of the thermal bath in which the memory is immersed, and the strength of the coupling to the bath. Since error correcting codes typically do not employ a Hamiltonian, we can consider their to be no correlations between the errors caused to each spin by the bath. This allows us to use a more fundamental quantity in place of the lifetime: the threshold probability that a errors occurs on the spins of the code.

When defining error thresholds for codes, we usually begin by assuming that the code is prepared perfectly in its stabilizer space. For the planar code, this means the anyonic vacuum. It is then coupled to a thermal bath, causing single spin errors to be applied randomly and independently to each spin with a certain rate. As an example, consider a bath that causes bit flips ($\sigma^x$ errors) to occur on each spin with a rate $\gamma$. The probability that a bit flip occurs within an interval $dt$ is then $\gamma dt$, and the probability that a net bit flip (that is an odd number of bit flips, since even numbers cancel) occurs within an interval $t$ is $p(t) = (1-e^{-\gamma t})/2$.

\begin{figure}[t]
\begin{center}
{\includegraphics[width=8.5cm]{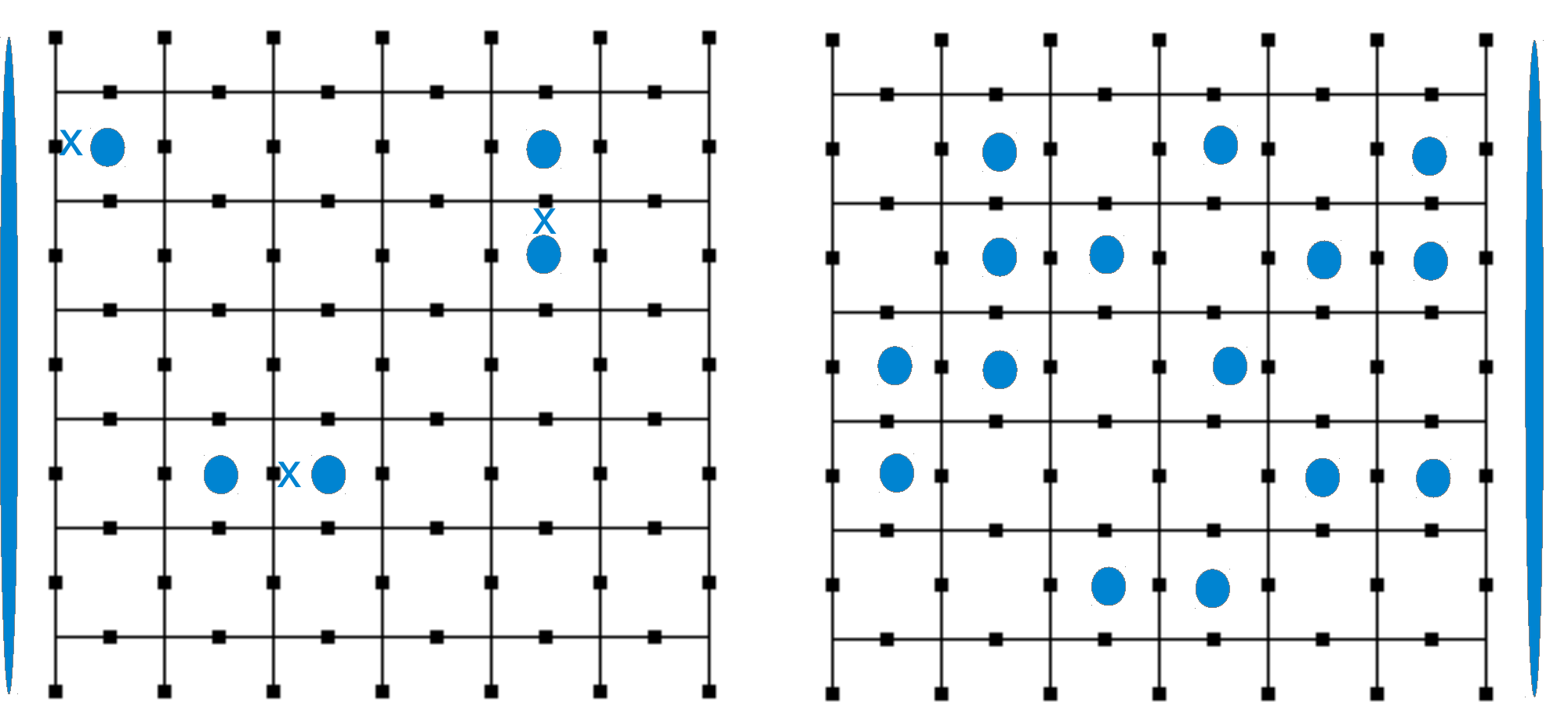}}
\caption{\label{threshpic} Example anyon configurations are shown for low (left) and high (right) rates for bit flip errors.  {In both cases, there was initially no $m$ anyon on either the left or right edges, meaning a logical $\ket{0}$ was stored. For the former configuration}, the pairing of anyons is simple, and it is easy to see on which spins the errors occurred. These are noted on the figure. It can been seen that the errors changed the occupancy of the left edge, and this can be corrected  {, returning the state to logical $\ket{0}$}. For the latter, the odd number of anyons in the bulk makes it clear that one of the edge occupancies has changed. However, the configuration does not make it easy to determine which one. If the attempt at error correction chooses the wrong one, the result will be to  {place an $m$ on both edges, leaving the state in logical $\ket{1}$ and hence applying a logical error rather than correcting it.}
}
\end{center}
\end{figure}

For small $t$, the probability that an error occurs on any spin will be small. The errors therefore result in the creation of isolated pairs of anyons. If we can assume that the stabilizers of the code, and hence the anyon occupancies, can be measured perfectly, we can easily see where these pairs are and determine which spins suffered an error. The correction of the errors is therefore straightforward. After long $t$, however, the probability of error will be close to $1/2$. Close to half of plaquettes and vertices will therefore hold an anyon, and it will not be clear which spin errors occurred to cause the configuration of anyons. Correction then becomes impossible. For examples, see Fig. \ref{threshpic}.

It is known that, between these two extremes, we will find a threshold probability $p_c$. If the rate of errors is below this, then error correction is always possible and the stored state can be retrieved with unit probability as $L \rightarrow \infty$. For error rates above this, correction cannot be performed, and fidelity of the stored state will become minimal as $L \rightarrow \infty$. This threshold probability then translates to a threshold time, the lifetime of the code, until which the stored state can be retrieved reliably. If correction is performed periodically, with the delay between correction rounds less than the lifetime, the quantum memory can be sustained indefinitely.

The retrieval of a stored state requires the use of an error correction, or decoding, algorithm. This is a classical algorithm that takes the results of the stabilizer measurements and determines the best guess for what errors occurred. The most naive decoding algorithm is brute force, which considers every possible sequence of errors consistent with the stabilizer measurements and assesses their likelihood. The use of brute force will ensure that the success rate of the decoding is the best that is possible, and that the correction succeeds all the way up to the threshold error rate. Practically, however, brute force is highly inefficient. The time taken to run the computation scales as $\exp{O(L^2)}$, and will take longer than the age of the universe even for small codes. Alternative algorithms are therefore required, which correct up to a lower threshold error rate than the theoretical maximum but do so with an efficient run-time. For any error correcting code, and any given error model, we will therefore have multiple threshold error rates. One will be the theoretical maximum, obtainable by brute force, and others will be the thresholds obtainable by various decoding algorithms.

The simplest error model that can be applied to the planar code is that of independent bit ($\sigma^x$) and phase ($\sigma^z$) errors. Let us use $p$ to denote the probability of a bit flip on each spin, and $p'$ to denote the probability of a phase flip. For the planar code defined on a square lattice, the threshold probabilities are found to be $p_c = p'_c = 11 \%$ \cite{dennis}. Both error rates must be below this limit for the stored quantum state to be retrievable. For codes defined on other lattices, the thresholds will change, typically with one rising above $11 \%$ and the other falling below. For example, using a hexagonal lattice results in thresholds of $p = 6.74 \%$ and $p' = 16.4 \%$. For a triangular lattice these two numbers will be exchanged. A series of randomized lattices have been found to interpolate between these extremes. In all cases it was found that the threshold achieves the so-called hashing bound, meaning that the planar code is highly optimal \cite{beatrand}.  The lattice may therefore be defined to best suit the ratio between $p$ and $p'$, thus optimizing the lifetime. Practically, error correction may be performed using the minimum weight perfect matching algorithm \cite{mwpm, dennis, fowlerthres, fowlerthres2}. This typically achieves a threshold that falls short of the maximum only by around $0.5 \%$, and so is very effective.

Another well studied error model is that of depolarizing noise, which assumes that $\sigma^x$, $\sigma^y$ and $\sigma^z$ errors occur on each spin with equal rate $p/3$, where $p$ is the probability that an error of any sort occurs. In this case, the threshold has been found to be $p_c = 18.9 \%$ \cite{bombinthres}. Minimum weight matching is not so effective in this case, achieving only a threshold of $15.5 \%$. A renormalization based algorithm manages $16.5 \%$, but (theoretically) this can be increased by increasing the run-time \cite{renorm}. A Markov chain Monte Carlo algorithm has been found that achieves an $18.5 \%$ threshold, which should be possible to increase to the $18.9 \%$ maximum with greater computational cost \cite{woottonloss}.

The above error models assume that the stabilizers can be measured perfectly, giving an exact and noiseless account of the anyon configuration created by the errors. However, such an idealized case is not physically relevant. Practically, measurement of the stabilizers will also be subject to errors. This means that single shot readout of the memory is not possible, since a single noisy measurement of the stabilizers is insufficient to perform error correction. Instead, the stabilizers must be measured periodically, in order  {that} error correction can detect not only the spin errors that occur, but the measurement errors also.

In this spirit, an error model has been proposed and studied to give a realistic account of the noise that a system is subject to. This assumes the measurement scheme with CNOT's coupling to ancilla spins, as described above. Errors are assumed to occur on all elements of the system. Between measurements, depolarizing noise is assumed to act on all code spins with rate $p$. The preparation of ancillary spins into their initial state is assumed to fail with the same probability $p$. The CNOT gates are assumed to be noisy, with single spin depolarizing noise acting on both spins with rate $p$ and measurement is assumed to give the wrong answer with probability $p$. The current best algorithm, based on minimum weight perfect matching, gives a threshold probability of $p_c \approx 1\%$, the highest yet known for any local and two dimensional architecture \cite{fowlerthres, fowlerthres2}. For $p$ below this limit, and as $L$ and the number of measurements are made large, the stored information can be read out of the memory with arbitrarily high fidelity and will have an arbitrarily long lifetime. The theoretical maximum for the threshold probability is not known, but developments of the minimum weight perfect matching algorithm are constantly increasing the achieved threshold, and the generalization of the Markov chain Monte Carlo algorithm to the noisy measurement case is expected to yield a significant increase.

\subsection{Experiments and proposals}

The first experimental steps toward realization of the planar code were actually done for the toric code model, which is the same up to boundary conditions. In these, it was photons that were used for the `spins', with entangled states created by parametric down conversion. The entangled states created corresponded to those of the toric code for small system sizes. Nevertheless, the size was sufficient to allow anyons to be created and braided by rotations of the single photons, and for the non-trivial effects of the braiding to be measured \cite{ex1,ex2}.

The first of these experiments realized a single-plaquette and four-vertex code with four photons \cite{ex1}. A superposition of an $m$ anyon and vacuum was created on the plaquette using a rotation in the $\sigma^x$ basis on one of the spins. $\sigma^z$ operations were then used to create an $e$ pair on the vertices, move one of these around the plaquette and reannihilate them. This changed the relative sign of the $m$-vacuum superposition in a way that could be measured. The second experiment followed a similar procedure, but with a two-vertex and four-plaquette code realized with six photons \cite{ex2}.

A more complex eight photon set-up has recently been realized using an eight-photon cluster state  \cite{ex3}. This was used to demonstrate the error-correcting properties of the planar code. A single qubit was encoded and shown to be protected from single qubit errors, with a logical qubit error rate lower than tthat on the physical qubits.

Another route towards experimental realizations is that of cold atoms. A toolkit of methods that may be used to realize anyon models with optical lattices were explored in \cite{zoller}. Experiments concerning minimal instances of topological order have also been considered \cite{paredes,lang}.

One promising proposal for the full realization of the planar code, which would eventually allow its use as both a memory and a quantum computer, involves the use of quantum dots \cite{christoph}. The physical quantum memory in this case would correspond to the spin qubit of the dot. In order to create and manipulate the planar code states, some means must be found to perform the required two qubit gates. The traditional approach for this would be to use the tunnel coupling between the qubits \cite{davdan}. However, this requires the qubits to be placed very close together, making it difficult to fit in the required gates and wirings. A recent proposal allows this problem to be avoided through the use of a long range coupling, mediated by metallic floating gates \cite{luka}. It has been shown that these can be used to implement CNOT gates with high fidelity, allowing the stabilizer measurements to be made on a lattice of quantum dots, and the planar code state to be prepared and manipulated for quantum computation.

\section{Hamiltonian protection and self-correction}

Quantum error correcting codes are usually considered as a means to correct errors through measurement of the syndrome. However, one could also consider implementing an error-correction inspired Hamiltonian, whose terms reflect the syndrome measurement operators and whose spectrum energetically penalizes errors. Of course, this can only be realistically be considered for codes whose syndrome measurements are (quasi-)local, or such a Hamiltonian could never be implemented. The planar code (and the related toric code) is therefore a natural candidate for this approach, since the stabilizer operators of the are quasi-local, acting only on the spins clustered around each plaquette and vertex. A Hamiltonian built out of these operators may then be defined as follows,
\be
H = - J_s \sum_s A_s - J_p \sum_p B_p.
\ee
The ground state space of this Hamiltonian is the stabilizer space of the code, and the anyonic vacuum, in which the state of the quantum memory is stored. Anyons, which are created by errors, become quasiparticles whose creation requires an energy $2 J_s$ for $e$ anyons and $2 J_p$ for $m$ anyons. At zero temperature, it has been shown that this Hamiltonian is completely stable against (sufficiently weak) local perturbations. Their effects are exponentially suppressed with the linear size, $L$, of the code \cite{bravyihastings}. Th

 This is especially true when there is disorder present in the Hamiltonian couplings, which can induce Anderson localization of anyon transport \cite{woottonrand,stark,kay1,beatrand}.

Usually the code is simply considered as a quantum memory when the Hamiltonian is considered, with no hole-based quantum computation as described above. However, this is not always the case. Proposals for similar computation schemes have been made in the presence of the Hamiltonian, using its protection against local perturbations. These are based on the planar (or toric) code itself \cite{pachosabelian}, as well as higher dimensional generalizations with more complex anyonic behaviour \cite{mochon,wootton2,wootton3}. This then leads into the realms of topological quantum computation with non-Abelian anyons, which is the subject of other reviews \cite{notes,why,nayak}.

At non-zero temperature the Hamiltonian greatly increases the lifetime of the code, since creation of anyon pairs is suppressed by a factor $\exp [-O(\beta J) ]$, for $J = {\rm min}(J_s, J_p)$. This leads to a lifetime that scales as $\exp [O(\beta J) ]$. The exact form of the relation for the lifetime depends on the details of the system-bath coupling, the geometry of the lattice used for the code and other such issues. By taking such details into account, the code can be tailor made to ensure that the lifetime achieved is the highest possible \cite{abbas}.

Though this Hamiltonian gives a lifetime that increases exponentially with $\beta$ and $J$, it is not always practical to raise the coupling or lower the temperature. Realistic regimes may therefore have a lifetime that is far below what we desire for quantum computation. Ideally we would like the lifetime to diverge with a parameter that is easier to increase, such as the system size. Such memories are called `self-correcting' because the Hamiltonian must not only suppress the creation of errors, but also be able to undo them once they occur.

For a classical example, we may consider the Ising model. For a one-dimension chain of spins of length $L$, this has the Hamiltonian,
\be
H_1 = J \sum_i \sigma^z_i \sigma^z_{i+1}.
\ee
An energy cost $2J$ is given to the walls between domains of spin up and spin down particles, meaning the states with all spins up and all down are the degenerate ground states. Let us encode a classical bit in these states. The Hamiltonian is then an energetic means the enforce the repetition code. If coupled to a thermal bath, errors will cause the creation of pairs of domain walls which may then walk along the line, flipping spins.

 {To determine how well the Hamiltonian suppresses the effects of such errors, we can consider the energy barrier. This is the minimum amount of energy required to perform a logical operation (flipping all spins in this case), if the operation is performed one spin at a time. The minimum energy required to map from the logical state $0$ to the logical state $1$ is achieved if first a single spin is flipped to create two domain walls, and then neighbouring spins are flipped to move these along the line. The only energy cost is the initial $4J$ required to create the domain walls. The energy barrier between the two logical states is then $0(1)$ in this model.}

Consider instead the two-dimensional version of this model, on a square lattice with side-length $L$ and Hamiltonian,
\be
H_2 = J \sum_i \sigma^z_i \sigma^z_{i+1} + \sigma^z_i \sigma^z_{i+L}.
\ee
In this case the energy cost for a domain wall is not fixed, but instead scales linearly with the length of the boundary. The ground states are again those with either all spins up or all down, and so can be used to store $0$ and $1$, making this an alternative way to energetically enforce the repetition code. The main difference to the one-dimensional model is the mapping from $0$ to $1$, which requires the growth of a region with opposite orientation until the point that it takes up the majority of the lattice. This will require it to go via a state for which the  {length of the} domain wall, and hence the energy  {barrier}, is $O(L)$.

The different energy landscape of the two models causes radically different thermalization behaviour. The one-dimensional model has only a constant energy barrier for physical errors to become logical errors, whereas the two-dimensional model has a barrier which increases the system size. The latter is therefore far more likely to undo its errors than the former. This gives the one dimensional model a lifetime of $\exp [-O(\beta J) ]$, with no increase with system size. The two dimensional model, however, has a lifetime that increases exponentially with system size below the critical temperature (for the ferromagnetic phase transition) and polynomially above \cite{barrett}.

Going back to the quantum regime, it is easy to see that the planar code Hamiltonian is a direct parallel of the one-dimensional Ising model. An energy is required to create an anyon pair, but then no further energy cost is required for it to perform the walk around the entire system that causes a logical error. This $O(1)$ energy cost of a logical error can be seen as the reason for its lifetime that does not increase with system size \cite{alicki1,alicki2}. For a self-correcting quantum memory, which we would like to act in a way corresponding to the two-dimensional Ising model, a diverging energy barrier must be constructed  {to suppress the spurious application of both logical $X$ and logical $Z$ errors.}

The most obvious way to do this is to generalize the planar code to higher dimensions \cite{dennis}. It has been found that the three-dimensional toric code has a diverging energy cost of logical bit flip errors, but not for logical phase. It therefore serves as a self-correcting classical memory, but the lifetime of the quantum memory is $O(1)$ \cite{alicki3,claudios}. The four-dimensional planar code Hamiltonian does serve as a self-correcting quantum memory \cite{alicki3}, but the three-dimensional nature of our universe would make the experimental realization of such a model quite difficult.

The greatest hope for planar code based self-correction therefore lies in the two-dimensional model with the addition of long-range interactions. Multiple different forms of such interaction have been proposed and studied. In \cite{hamma} a logarithmic attractive potential between anyons was proposed, as was a physical mechanism to realize it through coupling the code to an additional bosonic field. This leads to a self-correcting memory below a certain finite temperature phase transition, with lifetime scaling as $O(L^{\beta})$ with system size. Above the critical temperature the memory will still be self-correcting, but with the lifetime scaling only as $O(L^{2})$ \cite{new_wootton}. This transition only occurs if the toric code is used. For the planar code, the lifetime will scale as $O(L^{2})$ for any temperature.

In \cite{chesi,beatrand,hutter} a repulsive potential between anyons with power law decay is considered. Though one might consider this repulsion to increase the rate of logical errors, it also means that the energy required to create a pair of anyons depends on the number of anyons already present. This leads to a vanishing small anyon density at any finite temperature. The scaling of lifetime with $L$ is at most quadratic, and depends on the exact form of the potential. It has been found that this interaction can be engineered by coupling the code spins to cavity modes \cite{fabio}.

In \cite{hutter} a similar repulsive potential is also considered, but this time with an anyon-vacuum, rather than anyon-anyon, interaction. If the state is prepared in the anyonic vacuum, this means that $O(L^2)$ vacuum plaquettes and vertices will oppose the creation of any anyons. This results in an $\exp [ -O(L^2)]$ suppression of anyon creation, and hence a lifetime that scales $\exp [ O(L^2)]$.  {This coupling may also be realized through coupling to cavity modes \cite{hutter}.}

 {Recently a novel model has been proposed, in which the two-dimensional toric code is coupled to a three-dimensional ferromagnet \cite{fabhut}. All interactions in this model are local and of bounded strength, and no couplings to non-local features, such as cavity modes, are required. The effect of the ferromagnet is to induce an enhanced suppression of anyon creation, with the energy required to create a pair scaling linearly with $L$. This linear energy barrier exponentially suppresses the creation of anyons, and so will lead to a lifetime of $O(\exp L)$.}


Further proposals for self-correcting quantum memories go beyond the planar code to completely different models. It has been shown that any local stabilizer Hamiltonian in two-dimensions \cite{kaycol,bravter} cannot support self-correction, and indeed local Hamiltonians in any dimension will have limitations \cite{kay2}. More complex approaches are therefore required. One possibility is that of engineered dissipation \cite{pasta}. Also the Bacon-Shor code has been proposed as a possible candidate in three-dimensions \cite{shor,bacon}, however it is not known whether this achieves self-correction.

Also in 3D are the cubic code \cite{haah1,haah2}  {and welded solid codes \cite{michnicki}. Both of these have the kind of energy barrier one would expect from a self-correcting memory. The minimum energy required to perform spin-by-spin logical operators is $O(\log L)$ in the former and $O(L^{2/3})$ in the latter. However, self-correction has been shown not to be present for the cubic code \cite{haah2} and the same is strongly suspected for the welded solid codes \cite{michnicki}. The problem here is one of entropy. Increasing system size increases the energy barrier, but also increases number of paths which errors may take. If the latter increases to fast with respect to the former, errors will be able to overwhelm the barrier and not self-correct.}

 {Even so, it can be expected that self-correcting properties will still be present to an extent. If the system size is small enough with respect to the temperature, the number of errors present at any time will also be small, and so the energy barrier will have a significant effect. We can therefore expect the lifetime of quantum memory stored in such codes to scale as $O(L^{\beta})$ up to an optimal system size $L^* = \exp[O(\beta)]$. For any given inverse temperature $\beta$, this means a code can be constructed with lifetime $\exp [O(\beta^2) ]$, by ensuring that it is of the right size. The lifetime does not diverge with system size, and so the model cannot be called self-correcting. However, the $\exp [O(\beta^2) ]$ scaling of the lifetime is much better than the $\exp [O(\beta) ]$ scaling of models with no self-correction properties. It is therefore called `partially self-correcting'. The cubic code has been shown to behave in such a way \cite{haah2}, opening up a new realm of possibilities for quantum memories.}

\subsection{Experimental progress}

The first experimental realization of the toric code Hamiltonian was achieved using Josephson junctions \cite{glad}. This provided a demonstration of the degenerate ground state used for the quantum memory. The theory on which this experiment was based has a straightforward extension to the quantum double models \cite{jj}. These are qudit generalizations of the toric code which can also be used to store quantum memories, and which have more complex anyons \cite{toric}.

In general the major difficulty in realizing the planar or toric code Hamiltonian is the many-body vertex and plaquette interactions. Realistically, these need to be obtained perturbatively from two-body interactions. Possible means by which this may be done are using Kitaev's honeycomb lattice model \cite{honey}, or perturbative gadgets \cite{perturb}.

Progress is also being made into simulations of the planar (and toric) code model with Rydberg atoms \cite{rydberg1}. Rather than realizing the Hamiltonian directly, these seek to simulate its dynamics using the Rydberg blockade \cite{rydberg2}. Such a simulation could provide a demonstration of how well the quantum memory stands up to dissipative noise.

\subsection{Outlook}
 
Already we have seen a great deal of experimental progress in the use of error correction and other such protection mechanisms to enhance quantum memories. Many physical systems have been tamed and set to work storing quantum states, and small instances of quantum error correction have been achieved.

Further progress, as always, requires a closing of the gap between theory and experiment,  {with} concessions on both sides. Theory often wishes for big systems, fast gates and low noise, and expects measurement to be no harder than applying a projection operator. But none of these are the case, and so advances in encodings and protocols must be made to increase error thresholds and allow for the experimental complexity of performing the required operations. To make this easier, experiments must continue their push to make things bigger, faster and cleaner.

For the case of self-correcting quantum memories, the theory is still relatively new and experiments are non-existent. Both matters need to be addressed. There remains theoretical work to be done in understanding what sorts of systems allow self-correction, and finding examples of realistic models that could realize self-correction in the lab. The most advanced work so far is that done with long-range anyonic interactions \cite{fabio,hutter,fabhut}. But it would be advantageous to have additional and independent approaches for comparison. A two-dimensional partially self-correcting memory with local interactions, for example, may be regarded as a favourable alternative despite its non-diverging lifetime with system size. In any case, theorists must always keep in mind that their definition of realistic, i.e. using only local interactions, is not always the same as that of experimentalists.

\subsection{Acknowledgements}

The author would like to thank Jonathan Busch, Adrian Hutter and Fabio Pedrocchi for critical reading of the manuscript, and the Swiss NF, NCCR Nano and NCCR QSIT for support.

\end{document}